\documentclass[twocolumn, 10pt]{article}
\usepackage{graphicx,amsmath,amssymb,subfigure}
\usepackage[latin9]{inputenc}
\usepackage{hyperref}
\begin{document}
\title{
From Protein Interactions to Functional Annotation: 
Graph Alignment in {\em Herpes}
}

\author{Michal Kol\'a\v{r}, Michael L\"assig, Johannes Berg \\
Institut f\"ur Theoretische Physik,
Universit\"at zu K\"oln \\ 
Z\"ulpicher Stra{\ss}e 77, 50937 K\"oln, Germany\\
}
\maketitle
 
\subsection*{Abstract} 
\noindent
Sequence alignment forms the basis of many methods for functional
annotation by phylogenetic comparison, but becomes unreliable in the
``twilight'' regions of high sequence divergence and short gene
length. Here we perform a cross-species comparison of two herpesviruses,
VZV and KSHV, with a hybrid method called {\em graph alignment}. The method is
based jointly on the similarity of protein interaction networks and on sequence similarity. In our alignment, we find open reading frames for which interaction similarity concurs with a low level of sequence similarity, thus confirming the evolutionary relationship. In addition, we find high levels of interaction similarity between open reading frames without any detectable sequence similarity. The functional predictions derived from this alignment are consistent with genomic position and gene expression data.

%%%%%%%%%%%%%%%%%%%%%%%%%%
\subsection*{Introduction}

With the advent of genome-wide functional data, cross-species
comparisons are no longer limited to sequence information. A classic
extension of sequence alignment is structural alignment, which has
been used to compare evolutionary distant RNAs ~\cite{a-ha-etal-2005} and
proteins conserved in structure rather 
than sequence~\cite{a-bo-sh-2003,a-zh-sk-2005}. Here we use protein 
interactions as evolutionary information beyond
sequence~\cite{a-ue-etal-2005}.
%~\cite{a-ke-etal-2003,ke-etal-2003,BergLaessig:06}.

We perform a cross-species analysis of two herpes viruses,
the varicella-zoster virus (VZV\footnote{VZV: varicella-zoster virus, KSHV: Kaposi's sarcoma-associated herpesvirus, VOCs: viral orthologous clusters~\cite{a-hi-up-2000}, Orf: open reading frame.}), causing chicken pox and shingles,
and the Kaposi's sarcoma-associated herpesvirus (KSHV), responsible
for cancer of the connective tissue. The two viruses have diverged
approximately $200$ million years ago. Their sequence dynamics is
characterised by a high rate of point mutations (at least an order of
magnitude faster than their host populations~\cite{a-mc-co-1994}) and a
high rate of gain and loss of genes (an order of magnitude higher than
the mutation rates of procaryotes~\cite{a-al-etal-2001}). 
As a result, the sequence
similarity between the two species is in the ``twilight'' region of
detection by alignment: homologous proteins have an amino acid
sequence identity of about $20\%$.  Moreover, many open reading frames
are only about $60$ amino acids long.

The protein interactions in both species have recently been measured
by a yeast two-hybrid screen~\cite{a-ue-etal-2005}.  Together with regulatory
couplings, protein interactions are believed to be an important source of 
phenotypic change, possibly more so than the overall change of coding
sequences~\cite{a-King:1975,a-BeltraoSerrano07}.
Protein interactions are encoded in
mutually matching binding domains. The evolutionary dynamics of these domains is
governed by different selection and hence, by different tempi than the
overall coding sequence. Moreover, amino acids relevant for binding are
difficult to localise from sequence data alone, and the sequence of a
domain may evolve considerably while its interaction is conserved.
Therefore, we treat the experimental interaction data as evolutionary
information independent of sequence data. However, these data are
noisy as well, not least due to the experimental difficulties of
high-throughput measurements.
 
\begin{figure}[t!]
\includegraphics*{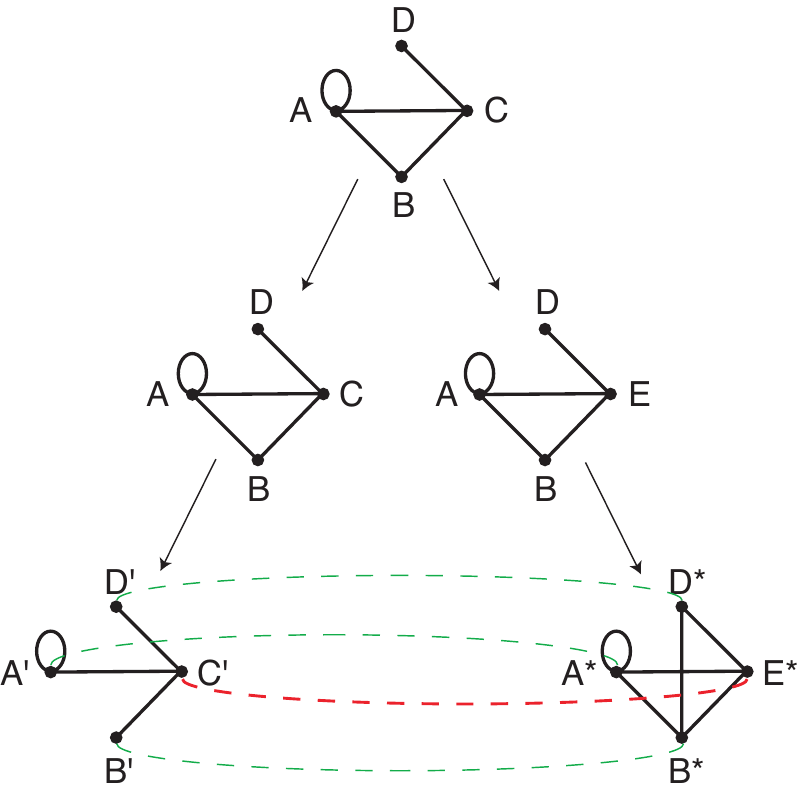}
\caption{\label{a-fig:nogd} \small {\bf Detecting functional relationships 
by graph alignment.}  In this example, the gene labelled $C$ is replaced in
  one lineage with its functional
  equivalent $E$, which has the same interaction partners in the
network. While some genes can still be correctly mapped across species
using sequence information (green lines), the full evolutionary
history and the mapping $C^{\prime}-E^{\star}$ are accessible from
cross-species analysis only by taking into account the interaction
networks.}
\end{figure}

Our hybrid comparison method called {\em graph alignment} jointly
uses the similarity of protein interactions and of coding
sequences to establish a mapping between genes of two
species~\cite{a-BergLaessig:06}. The underlying evolution involves a
number of distinct processes, including divergent sequence evolution,
gain and loss of interactions, duplication of genes and the
corresponding interactions, and gain and loss of genes. Functional relationships
may stem from common ancestry and thus be detectable by sequence {\em homology},
but they may also arise by convergent evolution, this {\em analogy} displayed by 
similar interactions without sequence similarity. 
An example is given in Figure~\ref{a-fig:nogd}, where one gene has
functionally replaced another gene by acquiring its interactions, a
process called non-orthologous gene
displacement~\cite{a-ko-mu-bo-1996}. Similarly, an
orthologous gene pair may diverge in sequence beyond detectability,
but conserved interaction patterns remain detectable due to functional
constraints. Such functional relationships are deduced from the network
of interactions between genes. Computationally, graph alignment rests
on a probabilistic scoring system, which weighs the cross-species
similarities of sequences and interaction networks based on their
evolutionary rates.  Our method simplifies in special cases: (i)~ If
gene sequences are well conserved and gene displacements are rare, one
may restrict the map between genes or proteins to sequence homology
(green lines in Figure~\ref{a-fig:nogd}) and, for example, identify 
network parts enriched in conserved links~\cite{a-SharanSuthrametal:2005,a-ba-sh-id-2006}. (ii) Conversely, if
sequence similarity has uniformly decayed below the significance
threshold, one can construct a map between genes or proteins based
solely on the interactions between them~\cite{a-Trusinaetal:2005}.

For the graph alignment between the VZV and KSHV viruses studied in
this paper, both the interaction networks and the gene sequences are
crucial to determine functional relationships, while each part of the data
by itself is often insufficient. In particular, we find protein pairs with
low sequence similarity for which the interaction similarity
strengthens the statistical inference of homology, as well as protein
pairs without sequence similarity, which are aligned based on their
interactions alone.  We use this alignment to make functional
predictions. These predictions turn out to be consistent with published experimental data
where available.

%%%%%%%%%%%%%%%%%%%%%
\subsection*{Theory}

\vspace{1ex} \noindent {\bf Scoring graph alignments.}
We use a recently developed method for the alignment of
graphs~\cite{a-BergLaessig:06}, adapted to the specific situation of
sparse interaction networks with a low number of matching links. Open
reading frames are represented by nodes, and pairwise protein
interactions are represented by links between nodes. A graph alignment
$\pi$ is a mapping of nodes of one network to nodes of the other
network. The alignment is characterised by  (i) interaction similarity of 
aligned nodes and (ii) sequence similarity of
the aligned nodes. 

Matching links in the two networks give a positive contribution to the
score: aligned node pairs $i$ and $j=\pi(i)$, and $i'$ and $j'=\pi(i')$
contribute a positive score if a link is present both between the pair
$(i,i')$ in one network and $(j,j')$ in the other network. An example
are the links between $D^{\prime}-C^{\prime}$ and
$D^{\star}-E^{\star}$ in Figure~\ref{a-fig:nogd}. A negative
contribution results if a link is present in one network, but not in
the other (mismatched links, as $D^{\prime}-B^{\prime}$ and
$D^{\star}-B^{\star}$ in Figure~\ref{a-fig:nogd}).  The values of the
link scores are encoded in the link scoring function $s_l(a,b)$ where
$a \in {1,0}$ (link present or absent in one network), and likewise $b
\in {1,0}$ (link present or absent in the other network). For binary
links this scoring function is a $2 \times 2$ matrix, conceptually
related to the scoring matrices in sequence alignment.
In addition to interaction similarity, the sequence similarities $\theta_{ij}$
between nodes contribute to the score, rewarding similarity 
between aligned pairs and penalising similarity between pairs not respected
by the alignment.

The total graph alignment score is the sum of independent
contributions from sequence similarity and from link similarity.  As a
result, a pair of nodes may be aligned because of high sequence
similarity, or because of high node similarity, or both. Of
course, the interplay between sequence similarity and link similarity
depends crucially on the relative weight of node score
and link score. These scoring functions are
determined self-consistently from the data within a Bayesian
framework, see supplementary text.

\vspace{1ex} \noindent {\bf Graph alignment algorithm.}
We use an iterative algorithm described in~\cite{a-BergLaessig:06} to
find the graph alignment with maximal score, based on a mapping to the
quadratic assignment problem. At each step the highest scoring
alignment is identified individually for each node, while keeping the
rest of the alignment fixed. A certain amount of
noise is used to help the alignment to escape from local score
maxima (as in simulated annealing~\cite{a-Kirckpatrick}).
%Kirkpatrick, S., C. D. Gelatt Jr., M. P. Vecchi, 
%"Optimization by Simulated Annealing",Science, 220, 4598, 671-680, 1983. 
This noise amplitude is gradually decreased to zero, starting
from some initial value $T$ and an initial alignment of reciprocal
best sequence matches.

\vspace{1ex} \noindent {\bf Alignment regimes and method tests.}
We test our alignment procedure on correlated random graphs, comparable in size and
average connectivity to the protein interaction networks of KSHV and VZV. 
Two such graphs are generated from a common ancestor graph by independently
adding and deleting a certain fraction of links (see supplementary text for 
details). In addition, we specify the sequence similarities for 
a subset of the node pairs. The correct alignment maps all orthologous pairs, including those where no sequence information is specified. 

For network pairs with high connectivity and high link similarity, the 
alignment is faithful and reproducible, independently of the noise level $T$.
For example, given $80$ nodes with $60$ sequence orthologs specified, 
and $140$ links of which approx. $100$ match, the alignment reproduces all nodes
with sequence orthologs and $70\%$ of those without. The finite recovery rate 
stems from node pairs lacking any matching links.  

For network pairs with low link similarity, we find two different 
alignment regimes depending on the initial noise level $T$. 

(i) In the {\em high-fidelity} regime for values of $T$ well below a threshold
value $T_D$, the alignment consists mainly of the nodes with sequence similarity, but does not extend much beyond. 
For example, if only approx. 50 of 140 links match in the above network pair,
the final alignment for $T=4$ contains all 32 node pairs with sequence similarity but only 4 node pairs without sequence similarity, all of which are correctly aligned. 

\begin{figure*}[t!]
\includegraphics*{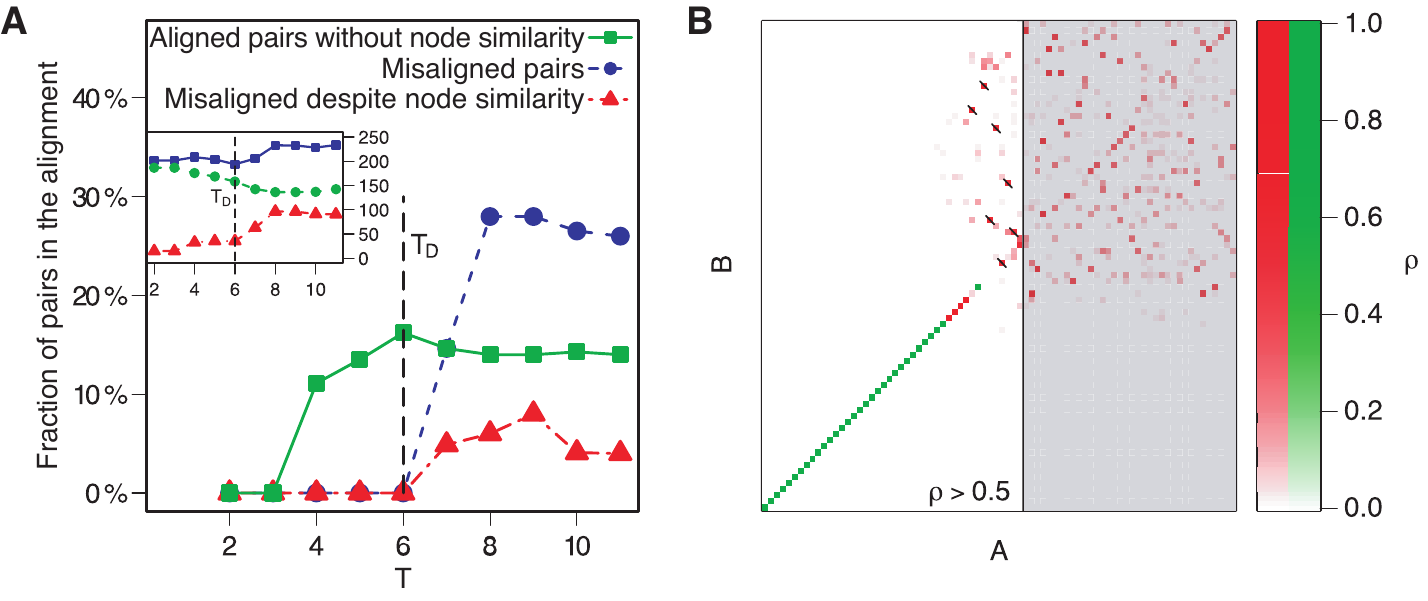}
\label{a-fig:weaksimER} 
\caption{\small {\bf Testing the graph alignment: Artificial networks
    with low link similarity.}
a) Unlike in the case of high link similarity (see text), the alignment 
of networks with low link similarity depends on the noise parameter
$T$ of the alignment algorithm.  With increasing noise parameter $T$ a
wider range of alignments is probed by the algorithm, leading to an
increasing number of nodes correctly aligned from their link similarity alone
(green $\square$-symbol). This heightened sensitivity is paid for with
decreasing specificity: also the fraction of nodes aligned with a node
different from their sequence homolog (red $\triangle$-symbols) and
the ratio of misaligned nodes without sequence homolog (blue
$\circ$-symbols) increase with $T$. Signature of the transition to 
the low-fidelity regime is the rapidly increasing link score 
(inset red $\triangle$-symbols) and the
decreasing node score (inset green $\circ$-symbols) while the total
score (inset blue $\square$-symbols) increases only slowly.  However,
just before the onset of the low-fidelity regime at $T=5$ correct
alignments are obtained from network similarity alone with few incorrect 
alignments (in this case none).  \newline
b) The noise parameter of the algorithm is
set just before the onset of the low-fidelity regime ($T=5$). The
alignment is represented by the matrix $\rho$, with $\rho(i,j)$ indicating
the relative frequency with which a node $i$ is aligned with $j$ over
many alignment runs. The correct alignment $i=j$ lies along the
diagonal. Entries of $\rho$ coloured green correspond to node pairs
with mutual sequence similarity, those coloured red have no sequence
similar partner and are aligned on the basis of link similarity alone.
A cutoff of $\rho>0.5$ is used, and aligned node pairs with less than
two matching links are discounted (crossed-out points). This leads to
$5$ node pairs correctly aligned on the basis of their links only, and
no misaligned node pairs. This is a conservative scheme, as can be
seen by glancing along the diagonal for additional entries with lower
values of $\rho$.}
\end{figure*}

(ii) In the {\em low-fidelity} regime for $T$ above $T_D$, high-scoring
alignments contain many link matches (even more than in the correct alignment),
but different runs have little overlap and most nodes (even with sequence similarity) are misaligned. Correspondingly, the alignment has a high link score and a low node score, see Figure~\ref{a-fig:weaksimER}(a). In the above example, only $29$ of $32$ nodes with sequence similarity are correctly aligned and 
 as many as $14$ of $21$ nodes without sequence similarity are misaligned at $T=8$. This behaviour is generic to graph alignment, independent of score
function and details of the algorithm. The high-scoring alignments 
in the low-fidelity regime are random islands of locally matching links.
Their occurrence can be understood from the special case of two uncorrelated graphs with a narrow
range of connectivities. Aligning a pair of randomly chosen nodes with
each other, their neighbours, and their next neighbours, etc., will
lead to a high link score (possibly offset to some extent by a low
node score). There are many such alignments with a high score, yet low
statistical significance. These spurious alignments occur for sparse
networks at sufficiently low fractions of link matches and low numbers
of nodes with sequence similarity. They are comparable to the score
islands known in local sequence alignment~\cite{a-hw-la-1996}, except
that in graph alignment the number of such ``islands'' is much greater
than in sequences.

(iii) Hence, {\em optimal detection} of similarity occurs in the  high-fidelity
regime for values of $T$ just below $T_D$; see Figure~\ref{a-fig:weaksimER}(b).
In this region the alignment is still guided by sequence similarity, yet
extends as much as possible into the set of nodes without sequence similarity.
The corresponding ROC curve is shown in the supplementary text. In the  above
example, this  conservative approach correctly aligns $5$ pairs of $48$ without
sequence similarity for $T = 5$, and there is no misalignment. The
same way of choosing $T$ is applied to the real protein interaction
network data.

%%%%%%%%%%%%%%%%%%%%%
\subsection*{Results}

\vspace{1ex} \noindent {\bf Optimal graph alignment between VZV and KSHV.}
The protein interaction network of the herpes virus VZV consists of $76$ Orfs and $173$
protein-protein interactions (of these Orfs, $19$ have no detected
interactions and are disregarded from the subsequent analysis). The
protein interaction network of KSHV consists of $84$ Orfs and $123$ interactions ($34$ Orfs
have no detected interactions). Thirty-four Orfs in VZV have
reciprocally best matching sequence homologs with reading frames in
KSHV. Between pairs of Orfs with such homologous partners, there are
$44$ interactions in VZV and $25$ interactions in KSHV. Of these
interactions, $8$ occur in both species, that is the overlap between
interaction networks is about $13\%$ when the alignment is given by
sequence homology.

The optimal alignment of the two networks is shown in
Figure~\ref{a-fig:alignKSHV_VZV}(a). For the list of aligned Orfs and details on the scoring see the supplementary text. The alignment consists of $26$ pairs of
aligned Orfs, spanning one third of the protein interaction networks of VZV and KSHV. The
alignment contains $44$ interactions, $10$ of which are
self-interactions. Of the $34$ interactions between distinct Orfs,
$11$ are matching interactions occurring in both protein interaction networks, only one of the
$10$ self-interactions matches. Of the $26$ pairs of aligned Orfs,
$24$ pairs have detectable sequence similarity. The remaining $2$
aligned pairs involve Orfs which have no detectable sequence
similarity with each other or any other Orf.  The mean connectivity of
the aligned part of the protein interaction network network is $3.0$ interactions per Orf,
compared with a mean connectivity of $2.4$ of VZV and $1.5$ of KSHV.

\begin{figure*}[t!]
\includegraphics*{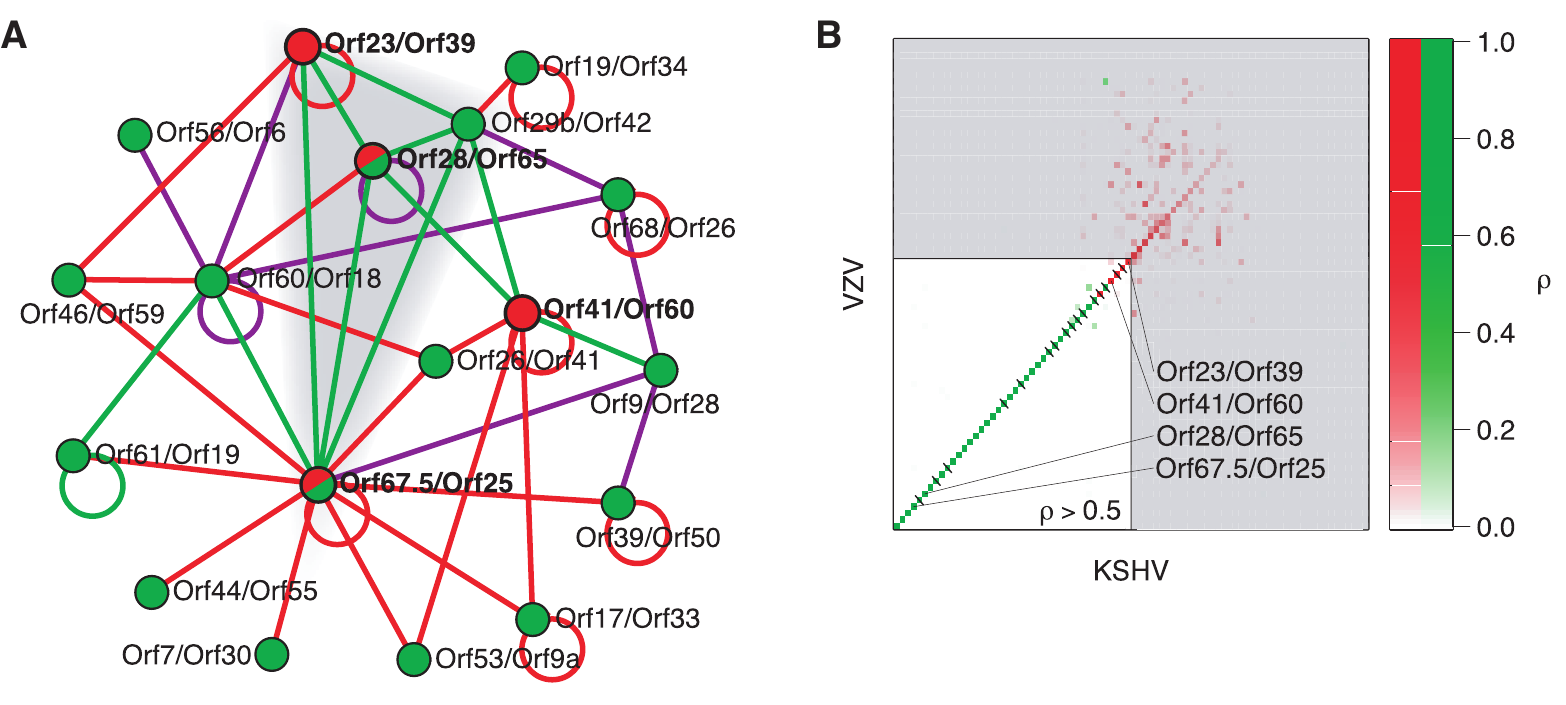}
\caption{\label{a-fig:alignKSHV_VZV} \small 
{\bf Alignment of the protein interaction networks of herpes viruses VZV and KSHV.}
a) The optimal alignment is shown with nodes representing aligned pairs of Orfs. 
Nodes are colour coded according to sequence similarity, 
measured by the sequence alignment score $\theta$ as described in the supplementary 
text. Green nodes have  
high sequence similarity with $\theta>0$, red nodes have no sequence
similarity detected, red/green nodes have low similarity with $\theta \leq 0$.
Protein interactions are represented by links between nodes, green
links indicate interactions which have been detected in both KSHV and
VZV. Interactions which have only been detected in KSHV or VZV are
shown in magenta or red, respectively. The cluster of matching
interactions linking nodes KSHV Orf23/VZV Orf39, 29b/42, 28/65,
and 67.5/25 is highlighted.\newline
b) The probing of the `twilight zone' of low and no sequence similarity by the alignment
is shown in the $\rho$-plot. Orf pairs with little or no sequence similarity are
aligned due to their matching interactions (red nodes, same colour 
scheme as in a).  The
conservative consensus alignment with $\rho>0.5$ is at the bottom
left. At lower values of $\rho$ spurious alignments occur (top right,
see Methods). The marked cases yield functional predictions discussed 
in the text.}
\end{figure*} 

\begin{figure*}[t!]
\vspace{1ex}
\includegraphics*{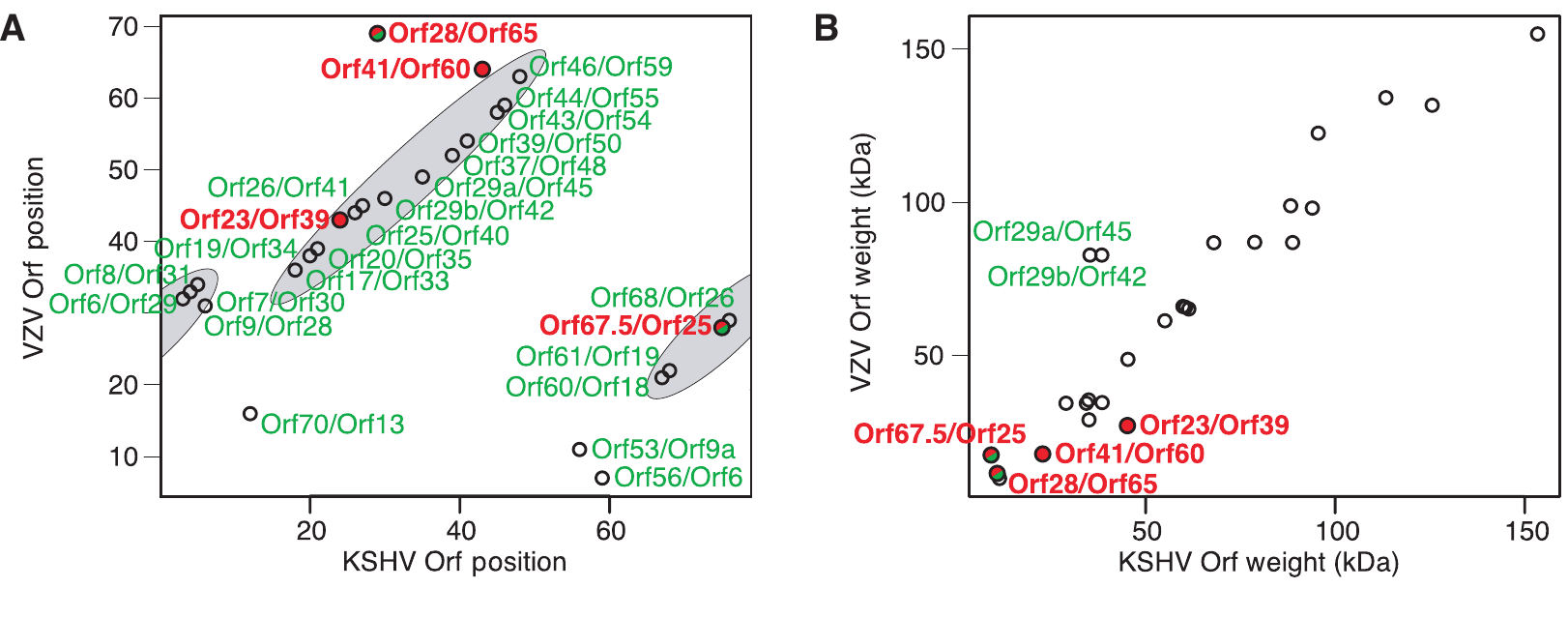}\label{a-fig:poshom}
\caption{\small \label{a-fig:corroboratingKSHV_VZV} {\bf Corroborating evidence
    for the network alignment from gene position and molecular
    weight.}  a) The gene rank of reading frames of VZV is plotted
against the rank in KSHV of their alignment partner. The points fall
into two diagonal bands indicating the conservation of gene order
between the two viruses. The Orf pairs aligned solely on the basis of
matching interactions fall within the those bands. The only
significant deviation from those bands, the pair KSHV Orf28/VZV Orf65, 
has related sequences, see text. \newline
b) The molecular weights of aligned pairs of reading frames
  show a strong correlation (Pearson's correlation coefficient
  $r=0.94$). The two exceptions again are aligned because they have
  related sequences (top left, indicated in green). The aligned Orfs
  with little or no sequence similarity (red circles, see text) show highly correlated 
  molecular weights.}
\end{figure*}

The quality of the alignment we have obtained can be tested by 
comparing the genomic positions of the aligned Orfs. We count  
the ranks of Orfs from the initial terminal repeats 
of the two genomes (left TR of KSHV, TRL of VZV).
In Figure~\ref{a-fig:poshom}(a) the ranks of reading frames in VZV are
plotted against the ranks of their alignment partners in KSHV.
Aligned Orfs without any sequence similarity fit very well 
into the sequence of Orfs in their respective genomes. 
In addition, the molecular weights of the aligned nodes 
are highly correlated, see Figure~\ref{a-fig:poshom}(b).

In some cases, sequence similar pairs of Orfs are not aligned because
of mismatched interactions. As an extreme case an Orf may have
several interactions in one species, but none in the other, indicating
most likely an unsuccessful Y2H experiment. Examples are KSHV
Orf64/VZV Orf22, 22/37, 42/53, 36/47, and 33/44.

\vspace{1ex} \noindent {\bf Functional relationships detected by interaction similarity.}
Some Orfs are aligned due to their matching
interactions, either with low or with no detectable sequence
similarity. We discuss these cases separately.

{\em KSHV Orf67.5/VZV Orf25.}
These Orfs have a sequence identity of only $18\%$ over $76$ aa (see
Methods for details).  They are listed as homologs in the VIDA3
database~\cite{a-al-etal-2001b}, and both of them are thought to be homologs of
the HHV-1 protein U$_\mathrm{L}$33~\cite{a-re-fa-ba-2000}. The alignment of
these Orfs largely results from $4$ matching links out of $5$ in KSHV
and $12$ in VZV (p-value of $4\times 10^{-3}$, see supplementary text
for details) with a local link score $S_L=4.57$ versus node score
$S_N=4.20$.  Our alignment thus confirms the homology.

{\em KSHV Orf28/VZV Orf65.}
These Orfs have a sequence identity of only $11\%$ over 102 aa.  They
are not listed as sequence homologs in databases
VOCS~\cite{a-hi-up-2000}, VIDA3~\cite{a-al-etal-2001b} and
NCBI~\cite{a-ba-etal-2004}. However, the sequence alignment extends 
over their complete length, with no gaps. 
Again, the alignment of these nodes results from $4$
matching links out of $4$ in KSHV and out of $5$ in VZV (p-value of
$10^{-3}$) with a local link score $S_L=6.30$ versus node
score $S_N=3.50$. Functional annotation is available only for VZV
Orf65; it belongs to the membrane/glycoprotein class, most
likely it is a type-II membrane protein~\cite{a-co-etal-2001}. The
alignment of KSHV Orf28 with VZV Orf65 leads us to predict that KSHV
Orf28 also codes for a membrane glycoprotein.

Several experimental studies support this prediction. Gene expression
studies show that Orf28 is co-expressed with tertiary lytic Orfs and
hence probably falls in the classes of structural or
host--virus-interaction
genes~\cite{a-je-etal-2001,a-mu-etal-2001}. The expression of
Orf28 is affected by blocking DNA replication~\cite{a-lu-etal-2004}
showing Orf28 is a secondary or tertiary gene. Furthermore, Orf28 has
been detected in the virion by mass spectroscopy, leading to a
tentative functional classification as a glycoprotein--envelope
protein~\cite{a-zh-etal-2005}. Finally, Orf28 is a positional homolog of
the Epstein-Barr virus Orf BDLF3, which is known to encode
glycoprotein gp150.

{\em KSHV Orf23/VZV Orf39.}
These Orfs have no significant sequence similarity: although the
alignment obtained with clustalW~\cite{a-clustalW} has a
sequence identity of $18\%$ over $240$ aa, it is statistically
insignificant; a randomised test yields a p-value of $0.43$. A
systematic analysis involving a wide range of different scoring
parameters does not yield a statistically significant sequence alignment either
(see supplementary text). The reading frames KSHV Orf23 and VZV Orf39
are aligned purely due to $3$ matching interactions out of $4$ of KSHV
and $4$ of VZV (p-value $2\times 10^{-2}$). The local link score
equals $4.47$ versus a node score of $-0.49$. Functional
classification is available only for VZV Orf39 as a
membrane/glycoprotein~\cite{a-al-etal-2001b}. The alignment thus leads
us to predict that KSHV Orf23 also codes for a membrane glycoprotein.

This prediction is supported by several experimental studies. Again
Orf23 is co-expressed with tertiary lytic Orfs~\cite{a-je-etal-2001}
and is sensitive to blocked DNA replication~\cite{a-lu-etal-2004}, so it
is a late gene. The expression patterns of Orf23 are similar to those
of structural and packaging genes.

{\em KSHV Orf41/VZV Orf60.}
These Orfs have $3$ matching interactions out of $3$ in KSHV and $6$
in VZV ($p=2\times 10^{-2}$), but no significant sequence similarity
(The \textsl{clustalW} sequence alignment has identity of $12\%$ over
$160$ aa with p-value $0.94$).  They are aligned with a local link
score of $4.39$ versus a node score of $-0.49$. Both Orfs are
functionally annotated. KSHV Orf41 codes for a helicase/primase
associated factor~\cite{a-wu-etal-2001} and is not affected by blocking
DNA replication~\cite{a-lu-etal-2004}. On the other hand, VZV Orf60 codes
for the glycoprotein L~\cite{a-al-etal-2001b,a-ma-etal-2000}.  It may be that
either of them has a so-far unknown function, leading to the matching protein 
interactions. 
This idea finds support in \cite{a-je-etal-2001}, where the expression
maximum of Orf41 was found to come after the secondary lytic
phase. This is surprising because the transcript is needed already
during the secondary lytic phase (DNA replication). No other
DNA-replicating gene controlled by a different operon to KSHV Orf41
has an expression dynamics with this property.  Such a delay of the
maximum of expression may have two reasons: either the transcription
of the Orf41 is not controlled after its role is finished, or Orf41
indeed has a hitherto uncharacterised function in the tertiary lytic
phase, possibly a structural one.

We also note that Orf41 is specific to the class of
$\gamma$-herpesviruses, of which KSHV is a member. Analogously, Orf60
is $\alpha$-herpesvirus specific. It is possible that the homolog of
Orf41 in VZV and the homolog of Orf60 in KSHV were lost as a result of
either of these proteins acquiring a new function.  This would be an
example of non-orthologous gene
displacement~\cite{a-ko-mu-bo-1996}. 

\vspace{1ex} \noindent {\bf Interaction clusters.}
The alignment shown in the Figure~\ref{a-fig:alignKSHV_VZV}(a)
contains a cluster of genes all interacting with each other. This
cluster comprises the aligned pairs KSHV Orf23/VZV Orf39, 28/65,
29b/42, and 67.5/25 connected by matching links only. The p-value for
such a fully connected cluster (a clique) to emerge at random is
approximately $5 \times 10^{-11}$. The pair KSHV Orf41/VZV Orf60
discussed above is connected to this cluster by two matching links,
forming an almost fully connected cluster of $5$ Orfs pairs with $8$
of $10$ possible links present and matching. Surprisingly, while all the other
Orfs in the cluster code for structural proteins (virion assembly
and structure proteins), Orf41 of KSHV is annotated as a
helicase/primase associated factor, and hence a gene involved in DNA
replication. The association with structure-related genes may be
interpreted as a further evidence towards another function of Orf41 as a
structural Orf.

The individual species contain further clusters, but these are 
not conserved across species. The cluster
comprising Orfs 28, 29b, 41 and K10 in KSHV contains genes coding for
predicted virion proteins, virion assembly and host--virus interaction
proteins. Orfs 25, 19, 27, and 38 forming a fully connected cluster in
VZV code for proteins involved in virion assembly, nucleotide repair,
metabolism, and host--virus interaction.

%%%%%%%%%%%%%%%%%%%%%%%
\subsection*{Discussion}

\vspace{1ex} \noindent {\bf Graph alignment results from sequence and interaction similarity.}
Our alignment of Orfs in two different herpes viruses yields a
cross-species mapping between Orfs based jointly on the correlation
between amino acid sequences and on the correlation between their
protein interactions. This approach is distinct from searching for the
overrepresentation of matching interactions among sequence
homologs~\cite{a-ke-etal-2003}. It allows the identification of homology
in cases where sequence similarity between two Orfs has decayed to
statistically insignificant levels. The resolution of this `twilight
region' of sequence similarity by using the information on protein
interactions is particularly relevant for the case of short genes
(such as in the present application), or high levels of domain
shuffling. It also allows to detect functional analogs, proteins with 
similar interactions but without common ancestry.

\vspace{1ex} \noindent {\bf Functional predictions from interaction similarity.}
We find several cases of Orfs with no detectable sequence similarity which are 
aligned with each other solely on the basis of matching
interactions. There are different possible mechanisms generating this situation; 
(i) a pair of orthologous genes lose their sequence similarity, 
and (ii) a gene functionally substitutes
for another gene. The original gene may then be excised from the
genome without phenotypic effect. This process has been termed
non-orthologous gene displacement~\cite{a-ko-mu-bo-1996}.

In both of these cases, sequence information is insufficient for
functional prediction. Based on the alignment due to matching
interactions and on the annotation of one of the alignment partners,
we predict the function of several Orfs. These predictions are
supported by gene expression experiments and by the genomic position
of the Orfs.

\vspace{1ex} \noindent {\bf Functional cluster as conserved subgraph.}
The optimal alignment (Figure~\ref{a-fig:alignKSHV_VZV}(a)) contains a
cluster of $4$ Orfs whose products all interact with each other in
both viruses. All members of this cluster belong to a single
functional class; they are involved in virion formation and structure and code for tertiary lytic transcripts.

There are other fully connected clusters both in VZV and KSHV, but
none of them occur in \emph{both} viruses.  These clusters contain
proteins in different functional classes; one cluster in VZV contains
proteins involved in virion assembly, nucleotide repair, metabolism,
and host--virus interaction.

The guilt-by-association
scheme of assigning like functions to interacting
proteins~\cite{a-ol-2000}
%Oliver, S. Guilt-by-association goes global. Nature 403, 601-603 (2000)
would fail in these cases. Refinement of the principle to
guilt-by-conserved-association, where functional correlation is only
assumed for proteins with an interaction in \textit{both} species,
correctly describes the functional correlations in the above clusters.
Looking at the functions of interacting genes in a single species, the
functional classes are only correlated very weakly (mutual information
entropy of $0.006$ bits, see supplementary text). However, pairs of
proteins with conserved interactions are more likely to share the same
function (mutual information entropy of $0.107$ bits).

Guilt-by-conserved-association might go beyond the statistical
significance gained from filtering false positives by cross-species
comparison. Interactions between proteins of the same functional class
are $1.6$ times as likely to be conserved between VZV and KSHV than
interactions between proteins of different functions. Correspondingly,
the mutual information on links between homologous pairs of Orfs is
nearly ten times higher for Orfs of the same function than for Orfs of
different function. This points to a particular mode of evolution of
protein interactions, namely interactions between proteins of like
function changing more slowly than those between proteins of different
function. Multi-species alignments of interactions networks will
provide an opportunity to address evolutionary questions of this type
by tracing the dynamics of interactions along the phylogenetic tree.

\vspace{1ex} \noindent {\bf Data deposition:}
The protein interactions for KSHV strain BC-1 and VZV Oka-parental
were taken from the yeast two-hybrid screens (Y2H) of the Peter Uetz
lab~\cite{a-ue-etal-2005}. The sequences of the two herpesviruses were
downloaded from the VOCs database~\cite{a-hi-up-2000} and the NCBI
database~\cite{a-ba-etal-2004,a-da-sc-1986,a-ru-etal-1996}.

\noindent {\bf Accession numbers:} {\bf Genomes}: KSHV: Human herpesvirus 8 strain cell line BC-1 (VOCs genome ID 890); VZV: Human herpesvirus 3 strain Oka parental (VOCs genome ID 921). {\bf KSHV Orfs}: Orf 67.5: provided by Peter Uetz, sequence follows: "MEYASDQLLP RDMQILFPTI YCRLNAINYC QYLKTFLVQR AQPAACDHTL VLESKVDTVR QVLRKIVSTD AVFSEARARP"; Orf 28: Genbank accession NP\_572080.1; Orf 23: NP\_572075.1; Orf 41: NP\_572094.1; Orf 29b: NP\_572081.1. {\bf VZV Orfs}: Orf25: VOCs ID 59436; Orf65: 59475; Orf39: 59450; Orf60: 59470; Orf42: 59453.

\vspace{1ex} \noindent {\bf Acknowledgments:} The authors thank Peter Uetz for several fruitful 
discussions and making the interaction data available prior to publication, 
Gordon Brown and Derek Gatherer for discussions on the protein 
sequence alignment, and Maria Mar Alb\`a for providing 
functional information data. Funding from the DFG is acknowledged under 
grants SFB 680, SFB-TR12, and BE 2478/2-1. This research was supported in part 
by the National Science Foundation under Grant No. PHY05-51164

%%THE SUPPLEMENT%%%%%%%%%%%%%%%%%%%%%%%%%%%%
\onecolumn
\noindent {\LARGE \textsc{Supplemental Text}}

\section{Networks comparison}
\subsection{Protein interaction networks data}
The protein interaction data comes from the work of the group of Peter
Uetz, \cite{ue-etal-2005}, and is publicly available as the supplement
of the cited article.

\subsection{Network alignment}
\begin{figure*}
\begin{center}
\subfigure[]{\includegraphics[width=0.45\columnwidth]{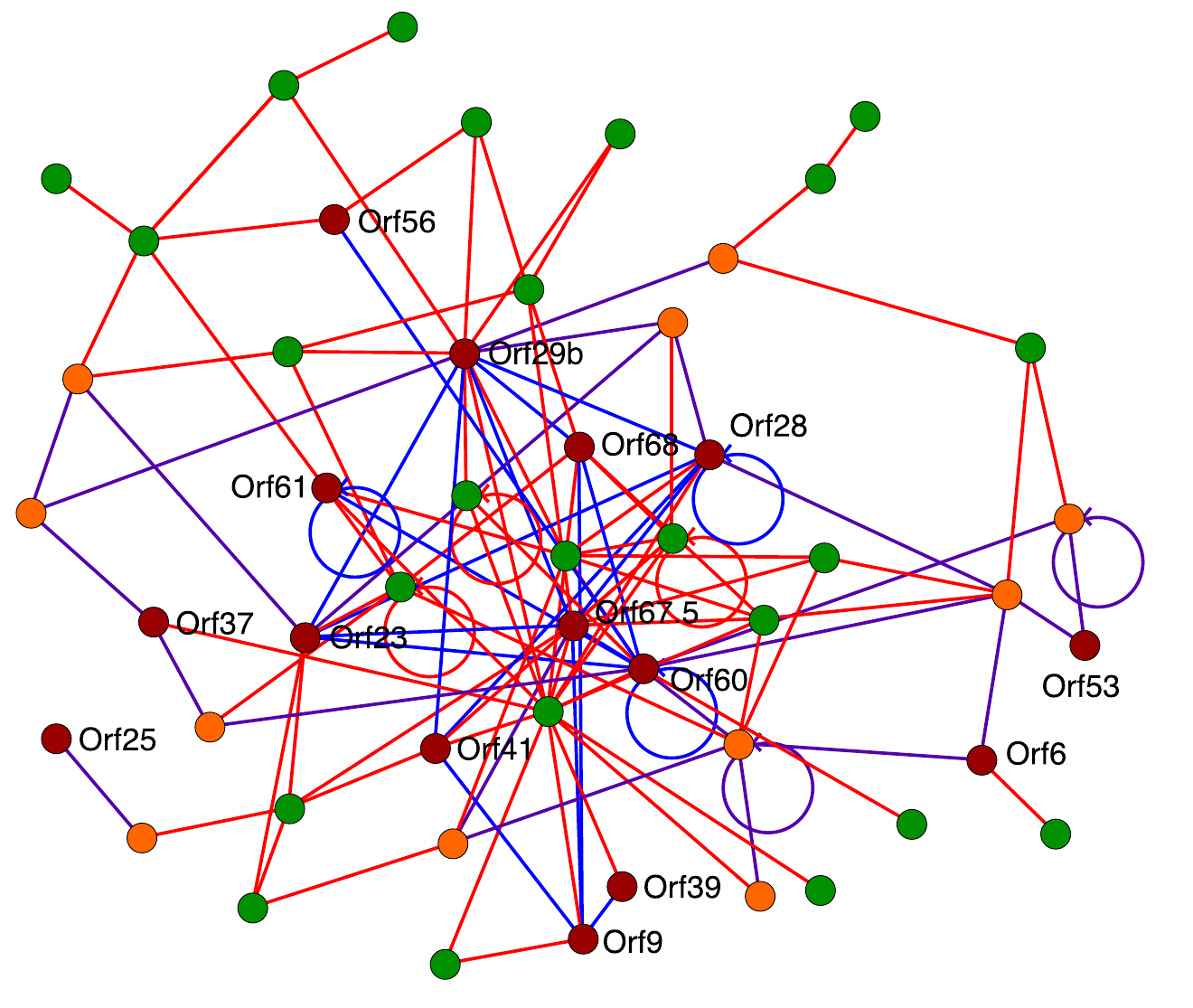}}
\subfigure[]{\includegraphics[width=0.45\columnwidth]{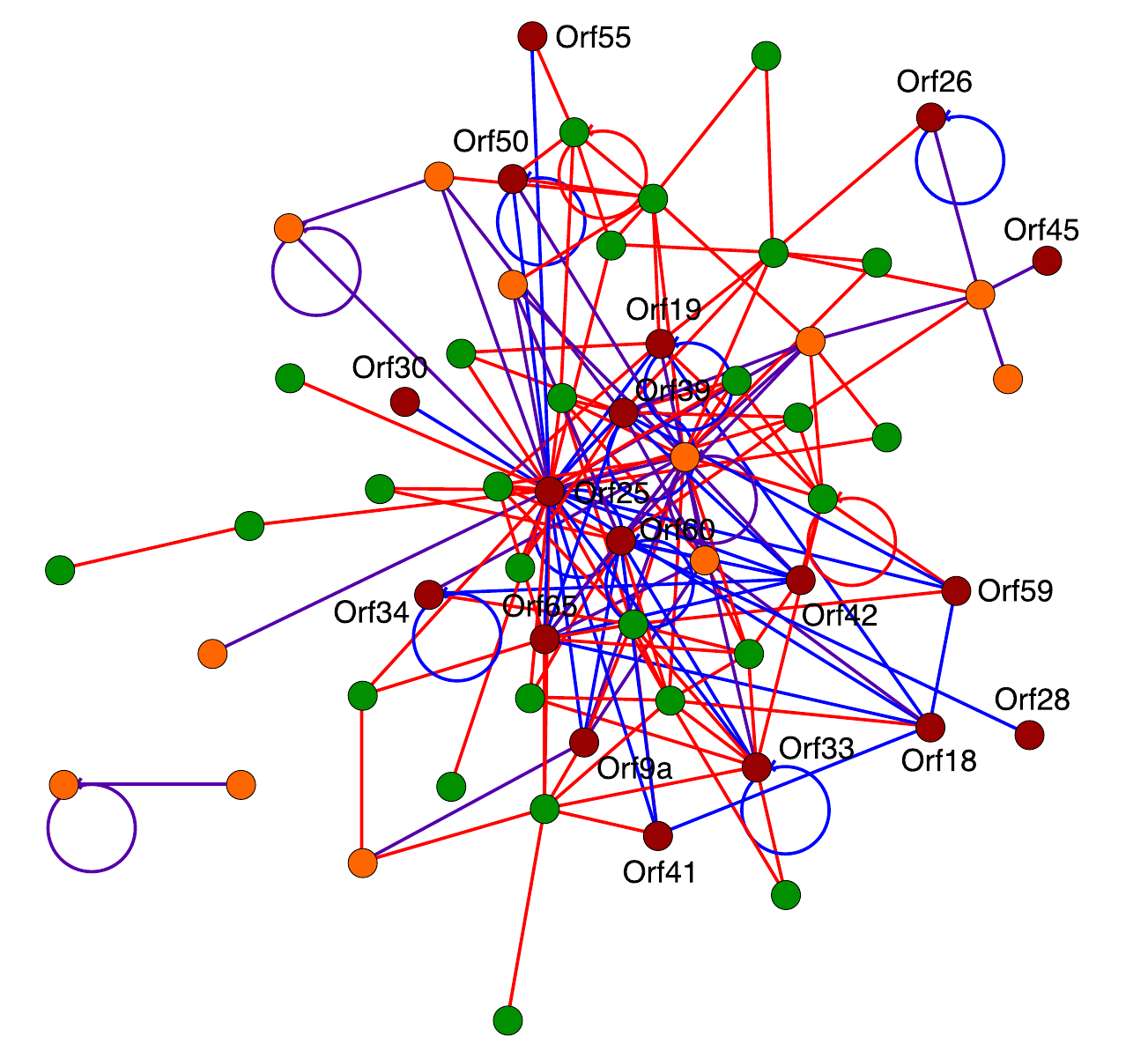}}
\caption{{\small {\bf The protein interaction networks of {\bf a)} KSHV and {\bf b)} VZV}. Each node in a network represents a single Orf. A link between two Orfs represents an observed protein--protein interaction. The subnetworks that belong to the final alignment are shown by dark red nodes and blue links. Their nodes are distributed homogeneously in the networks. The orange nodes are the Orfs that have been pruned out during the alignment pruning. Together with dark red nodes they show position of the consensus alignment. The violet links belong to the consensus alignment. All other nodes are plotted green and the links red.}}
\label{networks}
\end{center}
\end{figure*}

The protein interaction data are represented by a network (a graph) in
which each nodes represents an Orf of a species, and links denote experimentally observed
protein interactions. The two herpesviral protein interaction networks
are shown in the Figure~\ref{networks}\footnote{The experimental
  set-up allows orientation of the links by directing the links from
  the prey to the bait of the yeast--two--hybrid assay. In principle, one can use
  the resulting directed network for alignment. However, in that case the scoring
  matrix of the link score would has $6$ independent terms, as
  compared to $3$ for the undirected network, and the available data are not robust enough to
  infer their values.}.  We denote the KSHV%
\footnote{In the following text these abbreviations are used:
ER: 		Erd\H{o}s--R\'enyi (randomly generated);
KSHV: 	Kaposi's sarcoma associated herpesvirus;
Orf:		open reading frame;
PIN: 	protein interaction network;
VIDA:	VIDA virus database \cite{al-etal-2001b};
VOCs:	Viral Orthologous Clusters \cite{hi-up-2000};
VZV:		varicella--zoster virus.
}
network as $A$ and the VZV
network as $B$ when applicable. The networks $A$ and $B$ are described
by their adjacency matrices, which are square matrices with terms
$a_{ij}$, $b_{ij}$ equal $1$ if there is a link between Orfs $i$ and
$j$ in the respective interaction network and zero otherwise.

A \textit{network alignment} of two networks is a one-to-one mapping $\pi$ from the set of nodes (Orfs) of the network $A$ to the set of nodes of the network $B$: 
\begin{equation}
\pi: \{i \in A\} \to \{j \in B\}, \quad j = \pi(i).
\end{equation}
The nodes that are not aligned to any node in the other species, are
in our implementation aligned to a virtual dummy node.

Each network alignment may be assessed by a score combining both 
interaction and sequence data. The
score we define in the following text is a log-likelihood score
that stems from the comparison of two models: a model of evolutionary
related networks and the null model of independently created
nodes and links. The score has two parts;
the first contribution to the alignment score
quantifies the similarity of interactions of the aligned Orfs that is
the local topological likeness of the two networks. Hence it is
connected with the links of the networks and we term it the
\textit{link score} $S_L$. The other part utilises the sequence similarity of aligned Orfs, and hence it is connected with the nodes. We term it the \textit{node
score} $S_N$. The two contributions sum up to make the
total score of the alignment
\begin{equation}
\label{totalScore}
S = S_L + S_N.
\end{equation}	

\subsubsection{Link score}
The alignment of nodes induces an alignment of links; a link present 
between two nodes in one network may either be present or absent between 
their alignment partners in the other network. 
The topological part of the score is expressed as a sum of all
rewards for aligned links that are present in both networks
(\textit{matching links, conserved links}) and of all penalties for
the links that are present in one network only (\textit{mismatching
  links}). These rewards/penalties are parameterised by scoring
matrices $s_l$ for links between different nodes and $s_s$ for
self-links. If we denote by $A^\pi$ and $B^\pi$ the subnetworks of the
networks $A$ and $B$ that are aligned, that is the sets of nodes that
have alignment partners together with all links within these sets, we
may write the total link score as
\begin{equation}
\label{linkScore}
 S_L(\pi) =\sum_{(ij)\in A^\pi} s_l(a_{ij}, b_{\pi(i)\pi(j)}) + \sum_{i\in A^\pi} s_s(a_{ii}, b_{\pi(i)\pi(i)}) \ ,
\end{equation}
where $\pi(i)$ is the alignment partner of the node $i \in A$ in the
network $B$.

The parameters $s_l$ and $s_s$ are inferred by comparison of the model of evolutionarily related networks and of the null hypothesis in which the networks evolved independently. While in the independently evolved networks the existence of links between nodes $i, i^\prime$ and $j = \pi(i), j^\prime = \pi(i^\prime)$ are uncorrelated, the existence of links between homologous genes in the two evolutionarily related networks will correlate. The extend of this correlation, which depends on the evolutionary distance of the two networks, specifies the magnitude of the scoring parameters $s_l$ and $s_s$. We infer their values from the available protein--interaction data in the Section~\ref{ScoreParameters}.

\subsubsection{Node score}
In general, we expect that evolutionary related Orfs have correlated
sequences and hence we want to reward the alignment of nodes with 
correlated sequences and penalise the alignment of pairs of nodes with
dissimilar sequences. The measure of the similarity of the sequences
is the score $\theta$ of the sequence alignment which will be thoroughly defined in the Section~\ref{sequencesComparison}. The node score is then parameterised   
by some function $s_1(\theta)$, which is to be inferred from the data and which 
we expect to be an increasing function of $\theta$. Similarly, we also want to penalise the existence of pairs of Orfs
that are not aligned but have similar sequences. We expect the scoring
function $s_2(\theta)$ which parameterises this contribution to the
node score. Again this function needs to be inferred from the data and is 
expected to be a decreasing function of the sequence similarity
$\theta$. 

Both functions $s_1$ and $s_2$ express the differences between the evolutionarily related networks and the model of independently evolved networks. In the evolutionary model, we expect that homologous genes, that is genes with high sequence similarity $\theta$, are aligned. In the model of unrelated networks, we expect on the contrary, that the potentially aligned genes are not similar in their sequences. We expect also that in the evolutionarily model no pair of Orfs that are not aligned has high sequence similarity. Thus when compared to the unrelated networks of the null model, the frequency of such highly sequence similar pairs must be lower. The parameter functions $s_1$ and $s_2$ gauge the difference of the evolutionarily and unrelated--networks model and are inferred from the sequence and protein interaction data in the Section~\ref{nodeScoreSection}. The total node score is expressed as
\begin{equation}
\label{nodeScore}
S_N(s_1, s_2, \pi) = \sum_{i \in A^\pi} s_1(\theta_{i\pi(i)}) + \sum_{\mathrm{others}} s_2(\theta_{jk}) \ ,
\end{equation}
where we first sum all the rewards/penalties for the aligned pairs and then we add the contributions of all the pairs that are not aligned to each other but at least one Orf of the pair is aligned to some partner. The two symbolic sums may be rewritten for any alignment $\pi$ as 
\begin{equation}
\label{nodeScore2}
S_N(s_1, s_2, \pi) = \sum_{i\in A^\pi} \left[ s_1(\theta_{i\pi(i)}) 
    	+ \sum_{j\in B\setminus \pi(i)} w^\pi_{ij} s_2(\theta_{ij}) 
	+ \sum_{j\in A\setminus i} w^\pi_{j\pi(i)} s_2(\theta_{j\pi(i)}) \right],
\end{equation}
where $A\setminus i$ is the set of all nodes in $A$ but $i$ and the
factor $w^\pi_{ij}$ prevents overcounting of score contributions. Its
value is $1$, when only one of $i$ and $j$ is aligned, and $0.5$ when
both nodes are aligned to different partners. 

\subsubsection{Matrix representation of the score and the difference algorithm}
The problem of finding the optimal network alignment is mapped to the quadratic 
assignment problem, which is solved iteratively by repeated solutions of the Linear Assignment Problem. 

Both contributions to the alignment score, link and node score, can be
expressed in a matrix form. The node score is encoded in the matrix
$M^\pi$ and the topological part of the score in the matrix $R^\pi$:
\begin{equation}
\label{totalScore1}
      S = S_N + S_L = \mathrm{Tr} \; \pi (M^\pi + R^\pi) \ ,
\end{equation}
with
\begin{equation}
	  M_{ij}^\pi = \frac{1}{2} \sum_{k\in A^\pi, k\neq j, \pi(k)\neq i} s_l(a_{jk}, b_{i\pi(k)}) + s_s(a_{jj}, b_{ii}) \ ,
\end{equation}
and
\begin{equation}
      R_{ij}^\pi = s_1(\theta_{ji}) + \sum_{k\in B\setminus i} w^\pi_{jk} s_2(\theta_{jk}) + \sum_{k\in A\setminus j} w^\pi_{ki} s_2(\theta_{ki}) \ .
\end{equation}
$\mathrm{Tr} M$ stands for the trace of matrix $M$. 
  
For the sake of algorithm performance, we use instead of the total
score its generalised derivative, that is the change of the score upon
addition or removal of a node pair $ji$ to the alignment. This
approach makes the algorithm more greedy during the initial phase, yet
keeps it exact in the final stage. The derivative as represented in a
matrix form reads
\begin{equation}
    \Delta M_{ij}^\pi = \sum_{k\in A^\pi, k\neq j, \pi(k)\neq i} s_l(a_{jk}, b_{i\pi(k)}) + s_s(a_{jj}, b_{ii}) 
\end{equation}
for the change of the node score and  
\begin{equation}
      \Delta R_{ij}^\pi = s_1(\theta_{ji}) + \sum_{m\in B\setminus B^\pi, m\neq i} s_2(\theta_{jm}) + \sum_{m\in A\setminus A^\pi, m\neq j} s_2(\theta_{mi}) \ .
\end{equation}
for the change of the link score. The iterative update of the alignment is then done according to the scoring matrix $\Delta M^\pi + \Delta R^\pi$, by repeatedly 
solving the linear--assignment problem instance,~\cite{ma-to-1987}.
\begin{equation}
\label{update2}
	\pi \leftarrow \mathrm{argmax}_{\pi^\prime} \mathrm{Tr} \; \pi^\prime  (\Delta M^\pi + \Delta R^\pi + T \chi \eta) 
\end{equation}
until convergence. A noise term has been added so the algorithm
can escape from local score maxima: $\eta$ is a random matrix with
terms drawn independently at each iteration from the normal
distribution with the mean 0 and the standard deviation 1. The
addition of this random matrix similar to the simulated annealing
method of statistical physics, \cite{ki-ge-ve-1983}. The amplitude of
the noise starts at an initial value $T$ and decreases continuously;
the schedule function $\chi$ is a linear function decreasing from 1 at
the beginning of the algorithm run to 0 at its end.  The temperature
$T$ specifies the depth valleys in the score landscape the algorithm
can overcome and hence extent of the space of alignments that is
sampled. The higher the temperature $T$, the larger the volume of the
space of alignments that is sampled.

A package called \textsl{GraphAlignment} implementing this algorithm under 
the \textsl{R-project} is available for download on \cite{GraphAlignment}.

\subsubsection{Score parameters}\label{ScoreParameters}
In order to find the score parameters, we evaluate the likelihood of
the scenario in which the two networks evolved from a common ancestor
and compare it to the likelihood of the null model of two
independently created ER networks.

The likelihood of an alignment $\pi$ of the interaction networks $A$ and $B$ 
reads
\begin{equation}
 P(\pi | \{A,B\} ) 
 = \frac{P(\{A,B\}  | \pi) P(\pi)}{P(\{A,B\} )} \ ,
\end{equation}
where, within the Viterbi approximation~\cite{vi-1967}, the prior $P(\{A,B\})$ consists of the two terms in comparison: the evolutionarily related model ($\pi$) and the null hypothesis ($R$). 
\begin{equation}
 P(\{A,B\} ) = P(\{A,B\}  | \pi) P(\pi) + P(\{A,B\}  | R) P(R) \ .
\end{equation}
The likelihood may be expressed in terms of a log-likelihood score $S$
in the form of the sum of the link and node score $S = S_L + S_N$,
(\ref{totalScore}),\begin{equation} P(\pi | \{A,B\} ) = \frac{1}{1 +
    e^{-S(\pi)}} \ .
\end{equation}
This score consists of two independent terms: The contribution
depending on the network topology in the conditional probabilities
$P(\{A,B\}|\pi)$ and $P(\{A,B\}|R)$, and the contribution of the
priors $P(\pi)$ and $P(R)$. The independence of the two terms allows
us to assign the topological score $S_L$ to the first one and the node
score $S_N$ to the latter term,
\begin{eqnarray}
\label{totalScore2}
S_L & = & \ln \frac{P(\{A,B\} | \pi)}{P(\{A,B\} | R)}, \\ \nonumber
S_N & = & \ln \frac{P(\pi)}{P(R)}.
\end{eqnarray}
By identifying (\ref{totalScore2}) with (\ref{linkScore},
\ref{nodeScore}) we can readily find the scoring matrices $s_s$ and
$s_l$ together with the scoring functions $s_1$ and $s_2$. Before
doing so, we introduce a new quantity, the density matrix, which
allows us to control closely the behaviour of the aligning algorithm.

\subsubsection{Density matrix $\rho$}
For the evaluation of the scoring parameters we accumulate the results
of several runs of the alignment algorithm.

To store this data, we define the density matrix $\rho$
in the following way. We count the number of times $m_{ij}$ a pair of nodes $i\in A$
and $j\in B$ were aligned in $M$ runs of the algorithm, and set the 
corresponding matrix term $\rho_{ij}= m_{ij} / M$. We can rewrite the definition in terms of the resulting
alignments $\pi^\alpha$,
\begin{equation}
\label{density}
 \rho_{ij} = \frac{1}{M} \sum_{\alpha = 1}^M \delta(\pi^\alpha(i), j) \ .
\end{equation}
By $\pi^\alpha(i)$ we denote the alignment partner of the node $i\in A$ in the graph $B$ in the run $\alpha$. 
The term $\rho_{ij}$ of the density matrix then approximates the probability of finding the pair $(ij)$ in the final alignment.
    
\subsubsection{Mean values of the scoring parameters}\label{nodeScoreSection}

For the evaluation of the link score matrices we count frequencies matched/mismatched links 
in the alignment. That is, for each
pair $(i,i^\prime) \in A$ and the alignment partners $j = \pi(i)$ and $j^\prime = \pi(i^\prime)$ the terms
$a_{ii^\prime}$ and $b_{jj^\prime}$ of the entries of the adjacency matrices are compared and the
frequency table is accordingly updated. We calculate the frequency
tables both for the links and the self-links:
\begin{eqnarray}
    q_l(a,b) & = &\frac{1}{N_l} \sum_{(i,j)\in A} 	\sum_{(k,l) \in B} \rho_{ik} \rho_{jl} 
    \delta({a_{ij}, a}) \delta({b_{kl}, b}), \\ \nonumber
    q_s(a,b) & = &\frac{1}{N_s} \sum_{i\in A} \sum_{k 	\in B} \rho_{ik} 
    \delta({a_{ii}, a}) \delta({b_{kk}, b}) \ ,
\end{eqnarray}
where $N_l$ and $N_s$ are the normalisation constants of the two distributions and $a,b \in \{0,1\}$.

If the two networks evolved independently, as it is assumed in the null model, we can marginalise the frequency tables and find the probabilities of having a link between two nodes in the graph $A$ or $B$,
\begin{eqnarray}
p_l^A(a) & = & \sum_{b = 0}^1 q_l(a, b), \\ \nonumber
p_l^B(b) & = & \sum_{a = 0}^1 q_l(a, b). \\
\end{eqnarray}
By the marginalisation of the self link distribution, we obtain $p_s^A$ and $p_s^B$.

Finally, we obtain the score parameters $s_l$ and $s_s$ by comparing the null and evolutionary model,
\begin{equation}
\label{linkScoring}
s_r(a,b) = \ln \frac{q_r(a,b)}{p_r^A(a) p_r^B(b)}, \quad r \in \{l, s\}.
\end{equation}

Similarly, the node score parameters are inferred from the sequence similarities
$\theta_{ij}$ and the current alignment. Three situations may occur
for a pair of Orfs $i\in A$ and $j \in B$. Either the two Orfs are
aligned in $\pi$, or they are aligned to some other partners, but not
to each other, or they are not aligned to any partner. These three
disjoint sets of pairs of Orfs define three ensembles for which we
evaluate frequencies of the sequence similarity $\theta$;
$d_1(\theta)$ for the aligned pairs, $d_2(\theta)$ for the second
ensemble, and $d_0(\theta)$ for the pairs of nodes that are not
aligned. We take the score $\theta$ as
defined in the Section~\ref{sequencesComparison} as the sequence
similarity measure. The three distributions of $\theta$ are
\begin{eqnarray}
      \label{meanField2}
      d_1(\theta) & = & \frac{1}{N_1} \sum_{i\in A} \sum_{j\in B} \rho_{ij} \theta_{ij}, \\
      d_2(\theta) & = & \frac{1}{N_2} \sum_{i\in A} \sum_{j\in B} 
      (1-\rho_{ij}) 
      \left[ 1 - \prod_{k\in A, k\neq i} (1-\rho_{kj}) \prod_{l\in B, l\neq j} (1-\rho_{il}) \right]
      \theta_{ij}, \\
      d_0(\theta) & = & \frac{1}{N_0} \sum_{i\in A} \sum_{j\in B} 
      (1-\rho_{ij}) 
      \prod_{k\in A, k\neq i} (1-\rho_{kj}) \prod_{l\in B, l\neq j} (1-\rho_{il})
      \theta_{ij} \ ,
\end{eqnarray}
where $N_0$, $N_1$, and $N_2$ are normalisation constants. In general,
we expect $d_1(\theta)$ to be an increasing function of $\theta$,
reflecting the fact that the aligned Orfs should have similar
functions. Indeed, many sequence--homologous pairs belong to this
set. The distribution $d_2(\theta)$ is, on the other hand, expected to
be a decreasing function of $\theta$, similarly to $d_0(\theta)$.

The distribution $d_0(\theta)$ of similarities of unaligned Orfs may
be considered as the background distribution of $\theta$, and is taken
as the distribution in the null model. The node scores $s_1$ and $s_2$
read
\begin{equation}
\label{nodeScoring}
      s_r(\theta) = \ln \frac{d_r(\theta)}{d_0(\theta)},\quad r \in \{1,2\}.
\end{equation}

\subsubsection{Consensus and pruned alignment}
From the $\rho$ matrix we extract the consensus alignment as the
alignment of Orfs that have the corresponding $\rho$-matrix term
larger than $0.5$. This conservative choice of the cut-off is further
discussed in the following section.
    
The consensus alignment is then pruned in order to remove marginally
aligned pairs. These we define as the pairs that have a negative sequence 
score and at the same time less than two matching interactions. 
This pruning removes spuriously aligned pairs with both
low sequence similarity and low topological match. 

\subsubsection{Estimate of the p-value of the network alignment}
To calculate the p-value of aligning two nodes $i\in A$ and $j\in B$,
we remove the pair $(ij)$ from the alignment and find the probability
of placing in the vacancy a pair of nodes with a topological match as
good or better than the match of the pair $(ij)$. These two nodes are
chosen from two ER networks with sizes and mean connectivities
identical to those of the KSHV and VZV networks. These two unrelated
ER graphs correspond to the null model of independently evolved
networks.

A pair of nodes has the same or better topological match whenever it
has the same or a larger number of matching links to other aligned pairs
or it has a smaller number of mismatching links. For the pair
$(ij)$ with $r$ matching links in the alignment graph (Figure 3a in
the main text) the p-value is 
defined as the probability of finding a
nodes pair with $r$ or more matching links and at most $n_A-r$
(resp. $n_B-r$) mismatching links, where $n_A$ ($n_B$) is the total number
of links adjacent to $i$ in $A^\pi$ ($j$ in $B^\pi$).

This probability is easily evaluated for uncorrelated networks using 
the multinomial distribution. For $p_A \ll 1$ and $p_B \ll 1$ it reads 
\begin{eqnarray}\label{pValue1}
& & \!\!\!\!\!\!\!\!\!\!\!\!\!\!
p(r, n_A, n_B, N^\pi)  =  \\ \nonumber 
& &
(N_A - N^\pi + 1) (N_B - N^\pi + 1)  \\ \nonumber
& &
\times \;\; \sum_{m_A=r}^{n_A} \sum_{m_B=r}^{n_B} \sum_{s=r}^{s = \min\{m_A, m_B\}} \begin{pmatrix} N^\pi - 1 \\ 
s, m_A - s, m_B - s, N^\pi - 1 - m_A - m_B + s \end{pmatrix}\\ \nonumber
& & 
\times \;\; [p_A p_B]^s [p_A (1-p_B)]^{m_A - s} [(1-p_A) p_B]^{m_B - s} [(1-p_A) (1-p_B)]^{N^\pi - 1 - m_A - m_B + s} 
\ ,
\end{eqnarray}
where $N^\pi$ is the size of the aligned subnetworks ($N^\pi = 26$),
and $p_A$ and $p_B$ are the link probabilities in the two ER graphs
which we estimate from the complete KSHV and VZV networks
respectively, giving $p_A = 0.0330$ and $p_B = 0.0561$. The individual
terms of equation (\ref{pValue1}) can be understood intuitively: first
we choose a node in the network $A\setminus A^\pi \cup i$ (one node
out of $N_A - N^\pi + 1$), and a partner node from the network
$B\setminus B^\pi \cup j$. Next we choose from the $N^\pi-1$ remaining
nodes in the alignment network $s$ nodes that are connected by
matching links with the probability $p_A p_B$, $m_A - s$ (resp. $m_B - s$)
nodes that are connected by links only in the KSHV (VZV) subnetwork
with appropriate probability, and the remaining nodes that are not
linked to the pair $(ij)$ in either subnetwork. Finally, we sum over
all possible choices of the nodes (the multinomial coefficient) and
over all options that are equally good or better than the actual
alignment of $(ij)$. The contribution from the self--links (which are
typically mismatching) is close but smaller than $1$ and is neglected
here. The result is then an upper bound of the p-value. The estimated
p-values for the pairs of Orfs discussed in the main text are listed
in the Table~\ref{alignedPairs2}.

Similarly, we estimate the p-value of finding in the alignment
networks a clique with $M_C$ pairs, out of which $M_O$ pairs are
sequence related, and which are connected by matching links only. We
calculate this p-value as the probability of finding such a clique and
of finding among the links adjacent to the vertices of the clique the
same number or more matching links and the same number or less of
links that are present in one PIN only. Denoting the pairs of the
clique $(i^aj^a)$, where $a\in \{1,2,\ldots,M_C - M_O\}$, the numbers
of the matching links $r^a$, and the total number of links adjacent to
$i^a$ ($j^a$) in KSHV (VZV) as $n_A^a$ ($n_B^a$), this 
p-value is
\begin{eqnarray}\label{pValue2}
& & \!\!\!\!\!\!\!\!\!\!\!\!\!\!
p(M_C, M_O,\{r^a\}, \{n_A^a\}, \{n_B^a\}, N^\pi) = \\ \nonumber
& & 
\begin{pmatrix} N_A - N^\pi + M_C - M_O \\ M_C - M_O \end{pmatrix} 
\begin{pmatrix} N_B - N^\pi + M_C - M_O \\ M_C - M_O \end{pmatrix}
(M_C - M_O)!  (p_A p_B)^{M_C \choose 2} \\ \nonumber
& &
\times
\prod_{a=1}^{M_C-M_O} 
\sum_{m_A=r^a}^{n_A^a} 
\sum_{m_B=r^a}^{n_B^a} \!\!\! 
\sum_{s=r^a}^{s = \min\{m_A, m_B\}} 
\begin{pmatrix} N^\pi - M_C \\
s, m_A - s, m_B - s, N^\pi - M_C - m_A - m_B + s \end{pmatrix} \\ \nonumber
& &
\times \;\; [p_A p_B]^s [p_A (1-p_B)]^{m_A - s} [(1-p_A) p_B]^{m_B - s} [(1-p_A) (1-p_B)]^{N^\pi - M_C - m_A - m_B + s}
\ .
\end{eqnarray}
The contribution of self links is again omitted and giving upper bound
to the p-value. The formula (\ref{pValue2}) reduces to (\ref{pValue1})
in the case of an isolated node (1-clique) in which case $M_C = 1, M_O
= 0$.

The p-value for the clique formed by the pairs $67.5/25$, $28/65$,
$29b/42$, $23/39$ given by (\ref{pValue2}) is $5\times 10^{-11}$. The
p-values of finding such a clique in the PINs of the two species can
be estimated similarly and they are $2\times 10^{-3}$ in KSHV and
$4\times 10^{-2}$ in VZV. The difference between the p-value inferred
from the aligned networks and the p-values estimated from the
single--species networks indicates the significance of the evolutionary
conservation of the clique.

\subsection{Test of the procedure on artificial data}
\label{artificialData}
We test the performance of the algorithm on artificially generated 
networks with topological characteristics similar to those of the
actual KSHV and VZV networks. Since the correct alignment is known for
the generated networks, we can assess the specificity and
selectivity of the graph alignment in these cases. 

\subsubsection{High similarity of graphs}
We align two highly similar ER networks generated in the following
way. An ER network is generated with $98$ links and $80$ nodes. A copy
of this network is made and another $42$ links are placed in each network
independently of each other. For $60$ nodes chosen at random in one of the 
networks, we assigned node similarity with their ``orthologs''. 

\begin{figure*}
\centering
\subfigure[]{\includegraphics[height=0.4\textwidth]{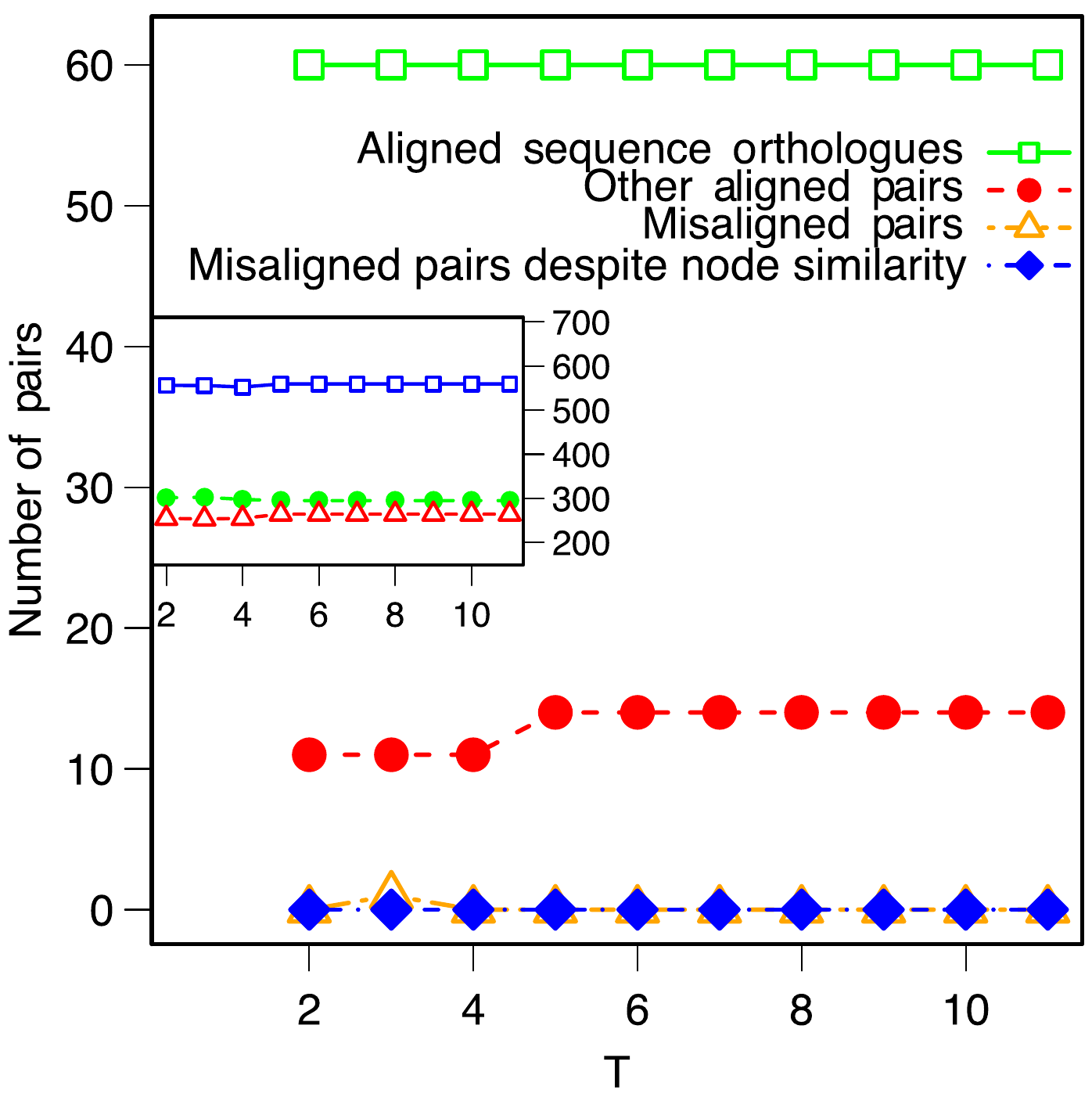}}
\hspace{0.017\textwidth}
\subfigure[]{\raisebox{1.5ex}{\includegraphics[height=0.39\textwidth]{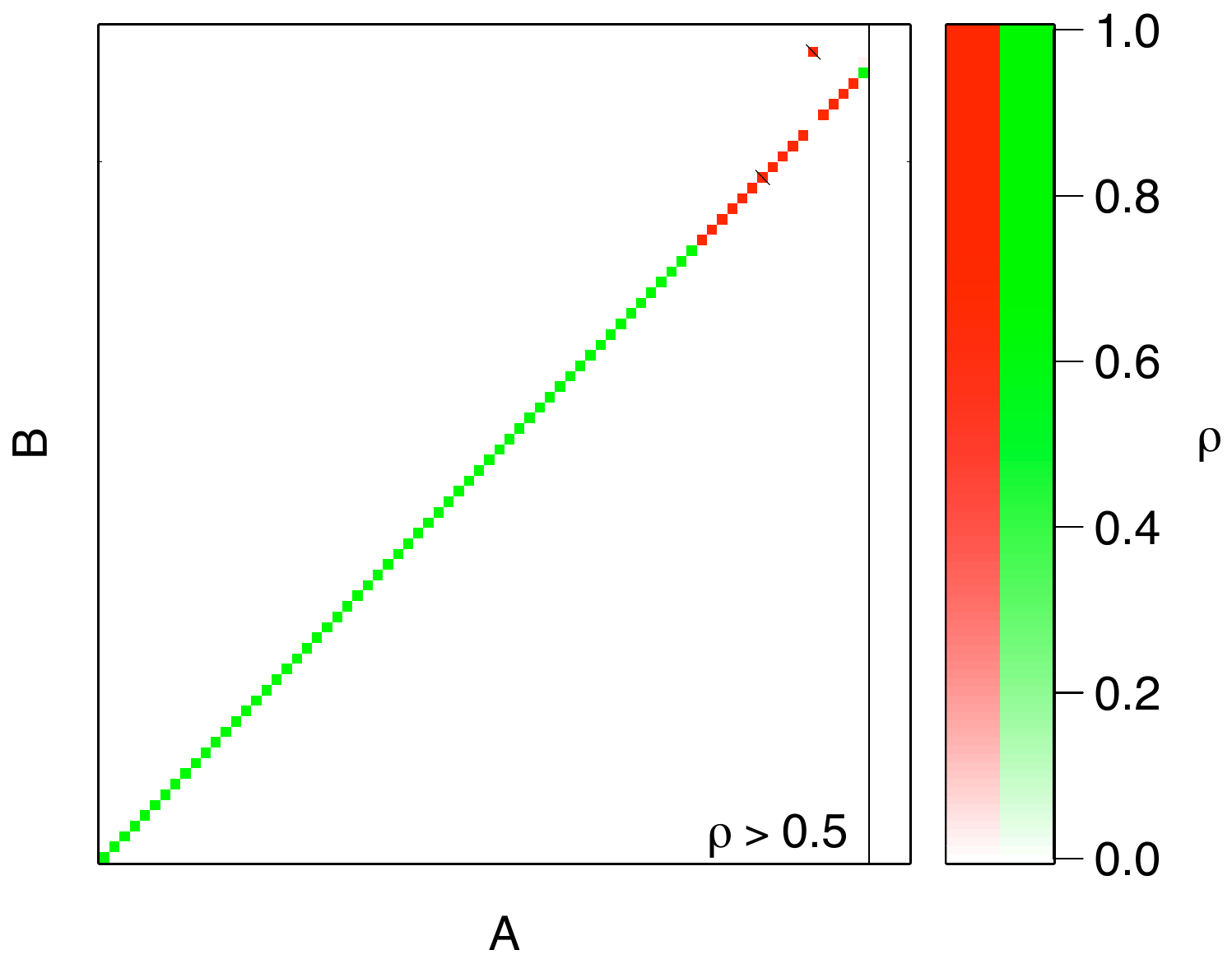}}}	
\subfigure[]{\includegraphics[height=0.4\textwidth]{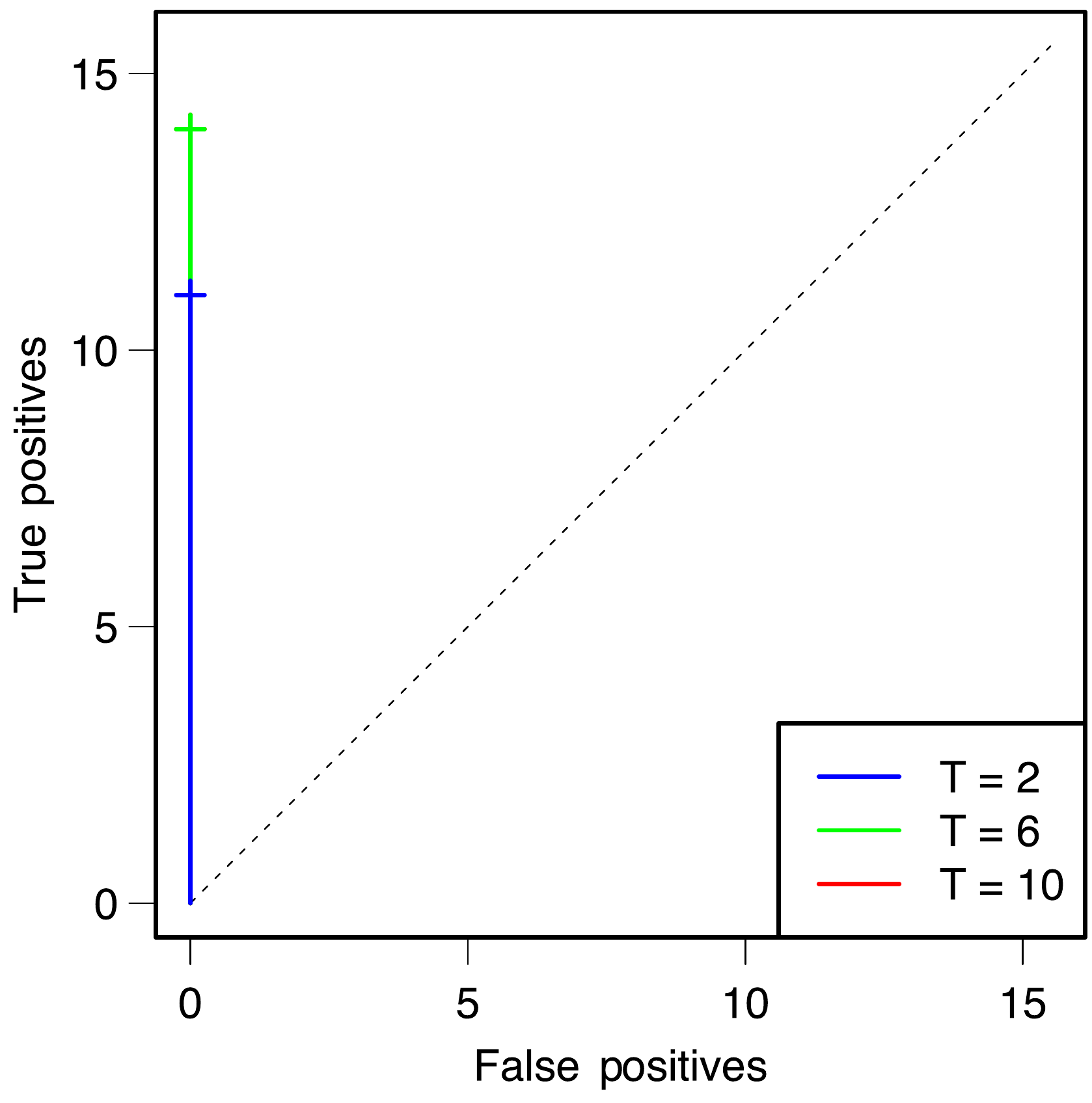}}
\caption{{\small {\bf Alignment of two highly similar Erd\H{o}s-R\'enyi
    graphs.}  \newline\indent\textbf{a) Main figure: Low noise level
    leads to sub-optimal alignments.} The dependence of the number of
  aligned sequence--homologous pairs ($\square$), the number of other
  aligned pairs ($\circ$), the number of wrongly aligned pairs
  ($\triangle$), and of the number of wrongly aligned pairs in which
  one or both partners are sequence homologous to a different Orf
  ($\diamond$) on the temperature $T$.\newline\indent{\bf a) Inset:
    For highly similar networks the score parameters are stable with
    updating.} The resulting total ($\square$), node ($\circ$), and
  link ($\triangle$) scores do not vary with the temperature.
  \newline\indent{\bf b) At $T=6$ is the alignment of the two random
    networks perfectly recovered.} The $\rho$ matrix gives the
  probability of aligning nodes pairs from the networks $A$ and
  $B$. The Orfs $i\in A$ are sorted with decreasing value of
  $\max_{j\in B} \rho_{ij}$. The Orfs of $B$ are on the other hand
  sorted in such a way that the diagonal of the density matrix
  corresponds to the correctly aligned pairs and the off-diagonal
  elements correspond to false positive predictions. The pairs of Orfs
  with detectable sequence similarity are shown in green, those with
  no sequence similarity in red. All pairs to the left of the vertical
  line denoting the $50\%$ cut-off belong to the consensus
  alignment. The pairs that have been pruned out in the pruned
  alignment are crossed out.  \newline\indent{\bf c) Another signature
  of the quality of the alignment are the numbers of true and false
  positives.} The numbers of true and false predictions are plotted
  and the cut-off value of $50\%$ is shown by the crosses. Only node
  pairs that do not have any sequence similarity are counted as true
  and false positives, hence the curve shows the result coming from
  the topological similarity only. The number of correctly aligned
  values increases with the temperature and reaches its optimal value
  at $T=6$.  }}
\label{figure}
\end{figure*}

The two resulting networks thus share 98 links out of 140 and contain
60 sequence--homologous Orfs out of 80. With such a high similarity we expect
the algorithm to find easily the correct alignment of the two
networks, which is also what we observe (93\% of the nodes are
correctly aligned, none is misaligned, see the
Figure~\ref{figure}). The quality of the alignment does not depend on
the details of the algorithm schedule, the updating of all scoring
parameters by the means of (\ref{linkScoring}) and (\ref{nodeScoring})
is possible. With increasing temperature, the number of aligned pairs
increases, yet the quality of the alignment is not compromised. The
optimal performance of the algorithm is reached at an intermediate
temperature $T=6$.

\subsubsection{Moderate similarity of graphs}
\begin{figure*}
\centering
\subfigure[]{\includegraphics[height=0.4\textwidth]{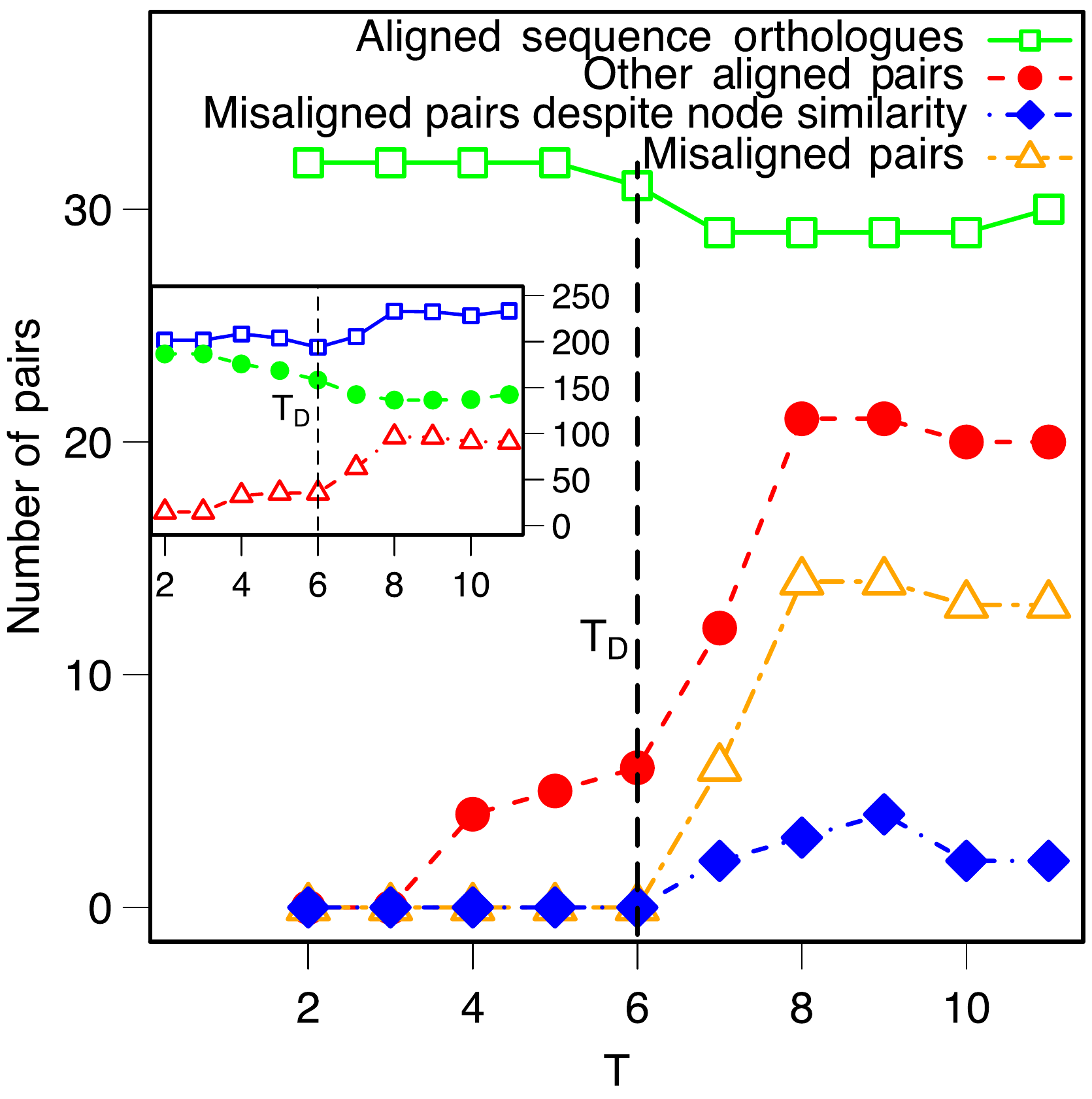}}
\hspace{0.17in}
\subfigure[]{\includegraphics[height=0.4\textwidth]{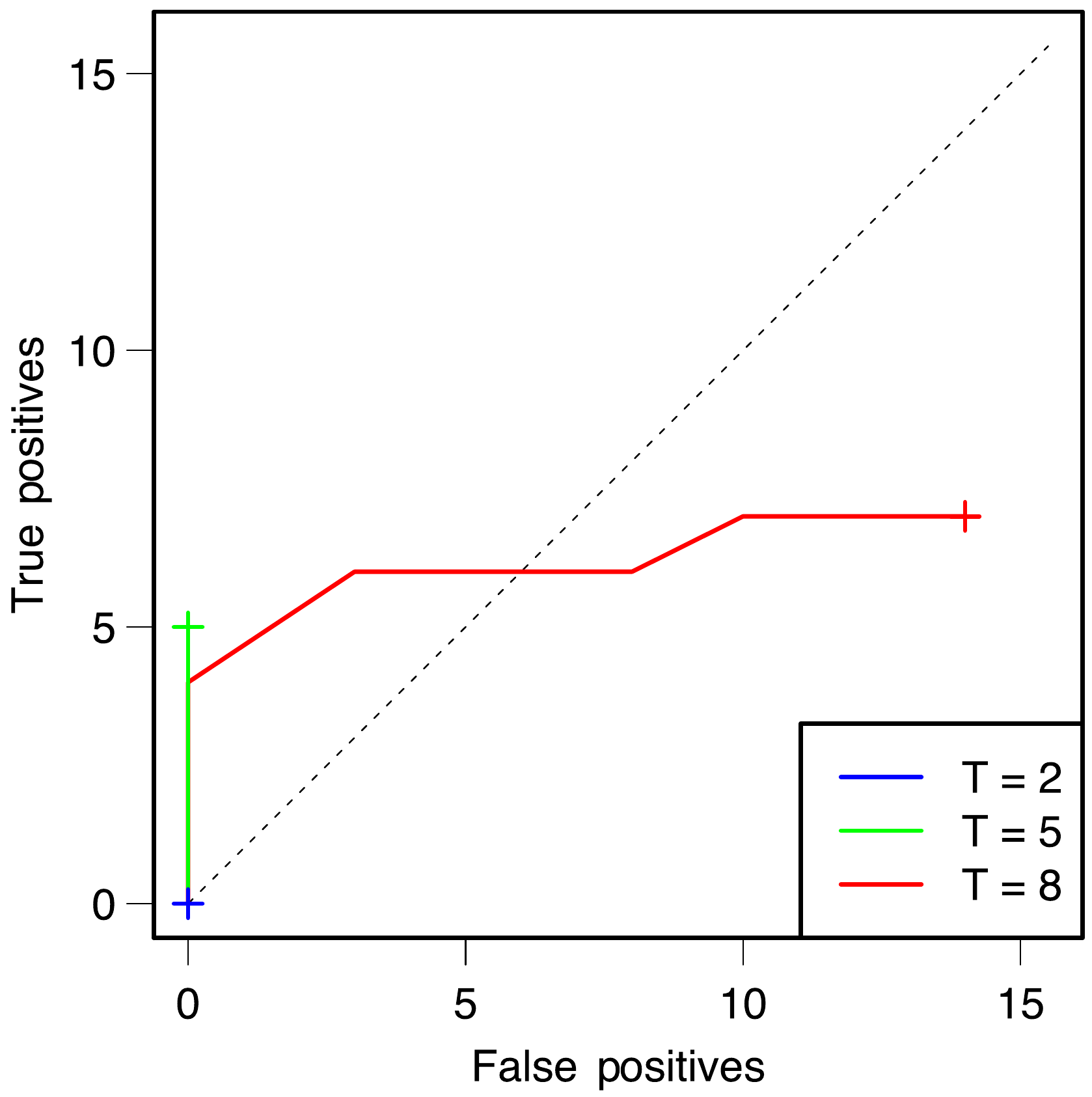}}
\caption{{\small {\bf Alignment of two moderately related Erd\H{o}s-R\'enyi
    graphs.}  For the legend see the caption of the
  Figure~\ref{figure}. See also the Figure~2 in the main text for the
  corresponding $\rho$-matrix.  \newline\indent{\bf a) Low-fidelity regime
    is linked to a steep increase of the link score.} Inset: The
  low-fidelity alignments have scores higher than the alignment of the
  sequence homologs, hence the search must be restricted only to the
  alignments in the vicinity of the sequence--homologs alignment. The
  temperature $T_D$ manifests itself in the steep increase of the link
  score.  \newline\indent{\bf b) Intermediate values of $T$ are optimal for
    searching the optimal alignment.} The false--true positives curves
  show that for too low $T$ the recovered alignment is trivial, and
  for too high $T$ it is faulty. The intermediate temperature $T=5,
  T\lesssim T_D$, shows the best ratio of true positives.}}
\label{figure2}
\end{figure*}

To test the performance of the algorithm in the regime appropriate to
the actual data, we repeat the test over graphs generated to resemble
the data. We generated a pair of ER graphs that out of $80$ nodes
and $140$ links share $49$ links and contain $32$ nodes with related
sequences, and hence their level of similarity is comparable to the
estimate for the real data. For these graphs we observe a nontrivial
dependence of the number of aligned and misaligned pairs on the
temperature; the choice of the temperature becomes crucial at this
level of network conservation. As the optimal temperature we have
chosen, in a conservative scheme, the largest temperature for which
the values of the score parameters remain close to their values
inferred from the initial alignment of sequence homologs, see
Figure~\ref{figure2}.

We observe that with decreased similarity of the two networks, the
optimal temperature at which we run the algorithm decreases to some
intermediate value. Already in the case of highly similar graphs we saw that 
the temperature must not be too low, for otherwise 
the sampled region of the alignment space is too
small. In the example of moderately related networks we see another
phenomenon: also too high a temperature decreases the quality of the
alignment. The case of  high temperatures is termed the low-fidelity regime, where 
the link score contribution grows quickly with temperature and
the node score, and hence the contribution from the sequence
similarity, decreases steeply. The essence of the phenomenon is best
described with a very simple example of aligning two Cayley
trees\footnote{The Cayley tree is a regular graph without loops in
  which every node has the same degree.}. If we distribute over the
trees ``sequence homologs'' densely, the correct alignment will be
recovered, as in the case of the highly similar Erd\H{o}s--R\'enyi
graphs. However, if we distribute the sequence homologs sparsely or we do
not place any homologs at all in the trees, the number of possible
alignments with perfectly matching links would be huge, but none of
these alignments would express the actual correlation of
the graphs. There would be no statistical significance
of the huge score. Updating of scores according to (\ref{linkScoring})
and (\ref{nodeScoring}) would result in a low or negligible node score
and a high link score, and consequently in a low reward for aligning
sequence related Orfs and an exaggerated reward for matching
links. This is exactly the behaviour that is observed in the
Figure~\ref{figure2}a.

To prevent the divergence of the scoring parameters we do not update
the link score parameters, instead we fix them to the values inferred
from the initial alignment of sequence homologs. Furthermore, being
rather conservative, we keep the temperature low, below the transition
value $T_D$, in order to restrict the search for alignments only
within the neighbourhood of the initial alignment in the configuration
space. By such a restriction, we recover some of the nodes pairs
properly even at low network similarity 
and at the same time we keep the false positives rate low,
see the Figure~\ref{figure2}b. At temperature $T=5$, that is slightly
below the critical temperature $T_D$, we correctly recover all $32$
sequence homologs, and another $5$ pairs of Orfs without sequence
similarity. In total, we recover $46\%$ of the complete network with
$80$ nodes. If we concentrate on the orthologs without sequence
similarity solely, the algorithm recovers at $T=5$ $10\%$ of
non-sequence homologous pairs. Here we note that among non-sequence
homologous pairs only $73\%$ have some topological similarity, $33\%$
share at least two links, and only $10\%$ share three or more
links. The aligned pairs without sequence similarity share two or
three links.

\section{Sequences comparison}\label{sequencesComparison}

\subsection{Genome data}
The sequences of the two herpesviruses (KSHV
strain BC-1 and VZV Oka-parental) have been downloaded from the VOCs
database~\cite{hi-up-2000}. Further Orfs (alternative
splices) have been obtained from the NCBI
database~\cite{ba-etal-2004} or the VIDA database~\cite{al-etal-2001b}
or have been provided by Peter Uetz~\cite{ue-etal-2005}. 

\subsection{Sequence alignment}
To assess mutual sequence similarity of the Orfs in the two viral
species we generate sequence alignments of each KSHV Orf with each VZV
Orf. Since the open reading frames are short and the level of sequence
similarity is low, care has to be taken in obtaining the optimal
alignment, as detailed below.

To account for the uneven level of sequence conservation across
the genome, we optimise the scoring parameters of the Needleman--Wunsch
algorithm individually for each pair of Orfs~\cite{ne-wu-1970}.  We
use affine gap penalties and the scoring matrices of the BLOSUM series
(BLOSUM35 to BLOSUM90, \cite{he-he-1992}).  We optimise the following 
parameters: the gap--opening penalty, the gap--extension penalty and
the evolutionary distance encoded by the BLOSUM matrices. The code for 
the sequence alignment, termed \textsl{sequenceAlign} is available upon request. 

\subsubsection{Score}
We define a standard log-likelihood score of an alignment of two
sequences by comparing a model based on evolutionary relation
of the two sequences with a random model. The random model of
independently evolved sequences depends only on the frequencies of amino-acids
occurring in natural peptides. If we denote these frequencies by
$p(a)$, where $a$ stands for an amino-acid residue and has $20$
possible values, we may write the probability of generating randomly a
sequence $\mathbf{a}$ of length $L$ with a composition $\{a_i\}$ as
\begin{equation}
P(\mathbf{a}) = \prod_{i = 1}^{L} p(a_i) \ .
\end{equation}
The probability of generating sequences $\mathbf{a}$ and $\mathbf{b}$ under 
this uncorrelated model reads
\begin{equation}
P^\prime(\lambda, \mathbf{a}, \mathbf{b}) = P(\mathbf{a})  P(\mathbf{b}) = \prod_{j = 1}^{L} p(a_j) p(b_j) \ .
\end{equation}
For evolutionary related sequences, we expect a higher probability observing two equal or similar residues, which is expressed by the log-likelihood score matrices $\sigma$. Hence
\begin{equation}
Q^\prime(\lambda, \mathbf{a}, \mathbf{b}) = \frac{1}{Z^\prime(\mathbf{a}, \mathbf{b})} \prod_{j = 1}^{L} p(a_j) p(b_j) e^{\sigma(a_j, b_j)} \ ,
\end{equation}
where $Z^\prime(\mathbf{a}, \mathbf{b})$ is a normalisation constant
\begin{equation}
Z^\prime(\mathbf{a}, \mathbf{b}) = \sum_\lambda \prod_{j = 1}^{L} p(a_j) p(b_j)e^{\sigma(a_j, b_j)} \ .
\end{equation}
The construction of the scoring matrices of the BLOSUM series
ascertains that the normalisation constant $Z^\prime$ equals $1$ for
sequences with the residue frequencies $p(a)$ close to those inferred
from current databases. This condition is also typically satisfied for
all proteins with $100$ and more residues. The log-likelihood score of
an alignment (without gaps) is then expressed as
\begin{eqnarray}
\theta^\prime(\lambda, \mathbf{a}, \mathbf{b}) & = & \ln \frac{Q^\prime(\lambda, \mathbf{a}, \mathbf{b})}{P^\prime(\lambda, \mathbf{a}, \mathbf{b})} \\ \nonumber
		  & = & \sum_{j = 1}^L \sigma(a_j, b_j) - \ln Z^\prime(\mathbf{a}, \mathbf{b}) \ .
\end{eqnarray}

With the proper normalisation of $Q^\prime$ by $Z^\prime$, the score
$\theta^\prime$ is larger than zero whenever the two sequences
$\mathbf{a}$ and $\mathbf{b}$ are more likely to evolve under the
evolutionary model underlying the scoring matrices in use.

To allow gaps in the global alignment we add two more parameters to
the model, the gap--opening penalty $\ln\mu$ and the gap--extension
penalty $\ln\nu$ (affine gaps). The score splits into two parts: the
substitutions score and the gap score:
\begin{eqnarray}
\theta(\lambda, \mathbf{a}, \mathbf{b}, \mu, \nu, \sigma) & = & \ln \frac{Q(\lambda, \mathbf{a}, \mathbf{b})}{P(\lambda, \mathbf{a}, \mathbf{b})} \\ 
\nonumber  & = & \sum_{\mathrm{aligned\;r.}\,j} \sigma(a_j, b_j) 
			+ \sum_{\mathrm{gaps}\,j} \left[\ln\mu + (l_j - 1) \ln\nu\right]
			- \ln Z^L \ .
\end{eqnarray}
Here we first sum all contributions from residue substitutions and
then we sum all the gap costs. The affine gap costs increase linearly
with the gap length $l_j$.

$Z^L$ is the normalisation constant of the probabilities $Q$ and it
depends on the length of the alignment $L$, the two sequences, the
scoring matrix in use, and the gap score parameters. Since the BLOSUM
score matrices are properly normalised by construction, $Z^\prime =
1$, or
\begin{equation}
\label{blosumNorm}
\sum_{a,b} p(a)p(b)e^{\sigma(a,b)} = 1 \ ,
\end{equation} 
the only contribution to $Z^L$ comes from the gaps. To calculate this
contribution, we will consider the following Markov chain.

We start with the two sequences completely unaligned and we choose one
option of: either \textit{(i)} we align the two initial residues of
the considered sequences (a substitution), or \textit{(ii)} we align
the initial residue of the second sequence with a gap, that is, we
create a gap on the first sequence (a deletion), or \textit{(iii)} we
create a gap on the other sequence (an insertion). In this way the
alignment is started and we extend it by one of the following steps:
either \textit{(i)} we align the residues that follow in the two
sequences (a substitution), or \textit{(ii)} we create the gap on the
first sequence (a deletion), or \textit{(iii)} we create a gap on the
other sequence (an insertion). We repeat the steps \textit{(i--iii)}
until the last residue is aligned to a residue or to a gap. The length
of the alignment $L$ is the number of the steps in the Markov chain.
For this Markov chain we can calculate the normalisation constant
$Z^L$ by a simple transfer matrix method. At each step $l$ there are
three possibilities of the end state of the alignment: either the last
step was a substitution, or a deletion or an insertion. Hence, we
split $Z^l$ in three parts $Z^l = Z_s^l + Z_d^l + Z_i^l$ that
correspond to the respective end-states. We may express the vector
$Z^{l+1} = (Z^{l+1}_s, Z^{l+1}_d, Z^{l+1}_i)$ at step $l+1$ of the
Markov chain as a function of the vector $Z^l$ at the step $l$:
\begin{equation}
Z^{l+1} = T Z^{l},
\end{equation} 
where the transfer matrix $T$ reads
\begin{equation}
T = 
\begin{pmatrix}
    1  & 1   & 1  \\
 \mu & \nu & 0 \\
 \mu & 0    &  \nu
\end{pmatrix}.
\end{equation}
At the beginning of the alignment process we may start with a
substitution, a deletion or an insertion and hence $Z^0 = (1, 1,
1)$. The normalisation constant for an alignment of length $L$ can be
readily calculated by applying the transfer matrix $L$-times on the
initial vector $Z^0$
\begin{equation}
Z^{L} = T^L Z^{0}.
\end{equation} 
For long alignments the dominant contribution comes from the largest
eigenvalue of the transfer matrix $\alpha$,
\begin{equation}
\alpha = \frac{\nu + 1 + \sqrt{(\nu-1)^2 + 8\mu}}{2} \ ,
\end{equation}
and it reads
\begin{equation}
Z^L = \frac{(2\mu - \nu + 3\alpha)}{\sqrt{(\nu-1)^2 + 8\mu}} \alpha^L \ .
\end{equation}
Since the logarithm of the normalisation constant $\ln Z^L = C(\mu,
\nu) + L \ln\alpha$ is extensive in the length $L$ and since $L$ is
the sum of numbers of substitutions, deletions and insertions, the
normalisation can be implemented as a shift of scores:
\begin{eqnarray}
\theta(\lambda, \mathbf{a}, \mathbf{b}, \mu, \nu, \sigma) & = & \sum_{\mathrm{aligned\;r.}} (\sigma(a_j, b_j) - \ln \alpha) \\ \nonumber
			& + & \sum_{\mathrm{gaps}} [\ln\mu - \ln \alpha + (l_j - 1) (\ln\nu - \ln \alpha)]	
			- C(\mu, \nu). 
\end{eqnarray}

The score defined by the last formula is properly normalised for any
choice of scoring parameters $\mu, \nu$ and $\sigma$, whenever the
substitution scoring matrix is normalised according to
(\ref{blosumNorm}). The normalisation is done against all alignments of length $L$, what is an approximation of the exact normalisation evaluated by Yu and Hwa, \cite{yu-hw-2001}, who considered all possible alignments of the two sequences. However, this approximation allows to evaluate the normalisation constant explicitly (instead of the iterative formulae of~\cite{yu-hw-2001}) and is at the same time a very good estimate for sufficiently large negative gap penalties. The normalisation allows us to search for the optimal
parameters for an alignment of any two sequences $\mathbf{a}$ by maximising 
the score $\theta(\lambda, \mathbf{a}, \mathbf{b}, \mu, \nu, \sigma)$ over arguments: the alignment $\lambda$ and the parameters  $\mu, \nu, \sigma$. This
maximisation is performed iteratively by the code \textsl{sequenceAlign}.

The final score is computed by subtracting the contribution of leading
and trailing gaps.  All alignments which are either too short ($6$
residues and less) or contain too many gaps (a gap opening every
$6^{\mathrm{th}}$ residue on average), are disregarded as
insignificant. The final score is used as the measure of the
sequence similarity $\theta$ which is used, in completion to
interaction data, in the network alignment.  For the remaining
alignments we compute also the percent identity defined as the number
of identities in the alignment divided by the total number of
substitutions in the alignment. Knowing the optimal alignment and its
score for all pairs of nucleotide sequences, we search for the
reciprocally best matching Orfs in the two species, considered bona-fide 
\textit{sequence homologs}.

The number of sequence homologs in the KSHV/VZV genome is 34, that
is approximately $40\%$ of the Orfs of each species. The list of the
sequence homologs and parameters of their alignments are given in the
Table~\ref{seqOrth}, together with the scores calculated using
\textsl{clustalW} (version 1.81, default
parameters~\cite{clustalW}). For the four Orfs pairs discussed in the
Results section of the main text we have estimated also p-values of the
\textsl{clustalW} alignment and we present the data in the
Table~\ref{alignedPairs1}.

Here we define the p-value of the \textsl{clustalW} alignment as the
probability of obtaining an alignment of two random sequences with the
same or higher percent identity and with a comparable length ($\pm
10\%$) as the alignment of the real protein sequences. 
To generate the ensemble of random sequences ($1000$ pairs) we permute the real
sequences in a random fashion. In this way we keep the lengths and
base compositions of the two sequences, but we remove any sequence
relation. The random pairs are afterwards aligned with
\textsl{clustalW} and the p-value is estimated from the frequencies of
the percent identities.
 
\begin{table}
\centering 
{\small
\begin{tabular}{|p{0.45in}|p{0.45in}|p{0.3in}|p{0.9in}p{0.9in}|p{0.9in}p{0.9in}|}
\hline
{\bf KSHV Orf} & {\bf VZV \hspace{0.2in}Orf} & {\bf seq. orth.} & {\bf \textsl{sequenceAlign}} identity (\%) & \hspace{0.7in} score & {\bf \textsl{clustalW}} \hspace{0.2in} identity (\%)  & \hspace{0.7in} score\\
\hline
%data are stored in the file homologs.data in Results directory
9       &28     &*      &43.3   &467    	&39.8   &2155 \\
70      &13     &*      &63.7   &369    	&61.8   &1247 \\
44      &55     &*      &36.5   &293    	&34.1   &1431 \\
25      &40     &*      &29.9   &281    	&27.0   &1628 \\
61      &19     &*      &32.6   &174    	&31.2   &1035 \\
60      &18     &*      &37.4   &162    	&37.6   &676 \\
29b     &42     &*      &38.4   &148    	&37.8   &680 \\
8       &31     &*      &24.8   &145    	&24.3   &924 \\
29b     &45     &       &41.2   &123    	&35.3   &643 \\
46      &59     &*      &39.9   &113    	&43.4   &560 \\
43      &54     &*      &24.5   &89     	&24.5   &668 \\
6       &29     &*      &20.1   &81     	&20.6   &790 \\
56      &6      &*      &30.2   &51     	&23.3   &726 \\
7       &30     &*      &23.6   &48     	&22.4   &548 \\
68      &26     &*      &20.0   &34     	&20.7   &370 \\
29a     &45     &       &28.5   &34     	&25.1   &305 \\
39      &50     &*      &18.9   &32     	&18.6   &280 \\
37      &48     &*      &23.5   &32   	&19.8   &291 \\
20      &35     &*      &34.2   &22   	&23.0   &154 \\
17      &33     &*      &28.1   &22   	&21.9   &311 \\
19      &34     &*      &19.8   &10     	&18.1   &304 \\
53      &9a     &*      &23.8   &8    	&28.6   &76 \\
26      &41     &*      &14.5   &4    	&20.3   &169 \\
67.5    &25     &*      &20.0   &0    	&18.4   &57 \\
28      &65     &*      &9.9    &0   		&10.8   &-31 \\
53      &8.5    &       &20.8   &-1   	&26.4   &57 \\
K6      &1      &*      &10.6   &-1   	&11.6   &-26 \\
69      &27     &*      &21.7   &-1   	&16.4   &126 \\
28      &8.5    &       &20.0   &-1   	&14.7   &-41 \\
67      &24     &*      &14.0   &-3   	&14.3   &2 \\
30      &8.5    &       &16.2   &-4   	&20.8   &6 \\
72      &7      &*      &10.2   &-7   	&13.4   &-42 \\
52      &1      &       &13.1   &-8   	&18.5   &7 \\
65      &0      &*      &20.6   &-9   	&18.2   &-19 \\
38      &49     &*      &21.7   &-9   	&18.0   &11 \\
72      &35     &       &6.6    &-9   	&13.3   &-38 \\
28      &1      &       &11.9   &-10  	&14.7   &-29 \\
53      &0      &       &13.9   &-11  	&17.3   &31 \\
52      &46     &       &17.7   &-11  	&19.9   &27 \\
K6      &S/L    &       &11.7   &-11  	&20.0   &5 \\
30      &9a     &       &14.5   &-11  	&19.5   &20 \\
53      &65     &       &6.9    &-11  	&15.7   &-17 \\
67.5    &49     &       &7.6    &-12  	&11.7   &-31 \\
K5      &58     &*      &11.4   &-12  	&12.7   &-48 \\
67.5    &9a     &       &15.4   &-12  	&15.6   &-19 \\
38      &7      &       &11.7   &-12  	&29.5   &16 \\
K15     &65     &       &8.1    &-12  	&10.1   &-42 \\
16      &69     &*      &11.5   &-12  	&10.9   &-42 \\
16      &64     &*      &11.5   &-12  	&10.9   &-42 \\
67.5    &7      &       &13.9   &-13  	&25.0   &6 \\
K8      &23     &*      &10.7   &-13  	&13.9   &-18 \\ 
K4      &S/L    &       &12.9   &-13  	&22.3   &18 \\
K4      &9a     &       &4.7    &-13  	&19.2   &1 \\
53      &1      &       &4.7    &-14  	&14.2   &-28 \\
30      &57     &       &10.0   &-14  	&17.4   &-23 \\
74      &36     &*      &10.0   &-14  	&12.9   &-30 \\
55      &58     &       &9.1    &-14    	&9.6    &-93 \\
\hline
\end{tabular}
}
\caption{{\small {\bf The detected sequence homologs:} We list all the pairs of putative sequence homologs detected by \textsl{sequenceAlign} together with the score and the percent identity returned by the code. The score and the percent identity obtained with \textsl{clustalW} (version 1.81, standard parameters values) are also listed. The pairs that are considered putatively sequence homologous are marked by asterisk.}}\label{seqOrth}
\end{table}

\section{Graph alignment of VZV and KSHV}

\begin{figure*}[f]
\centering
\includegraphics[height=0.4\textwidth]{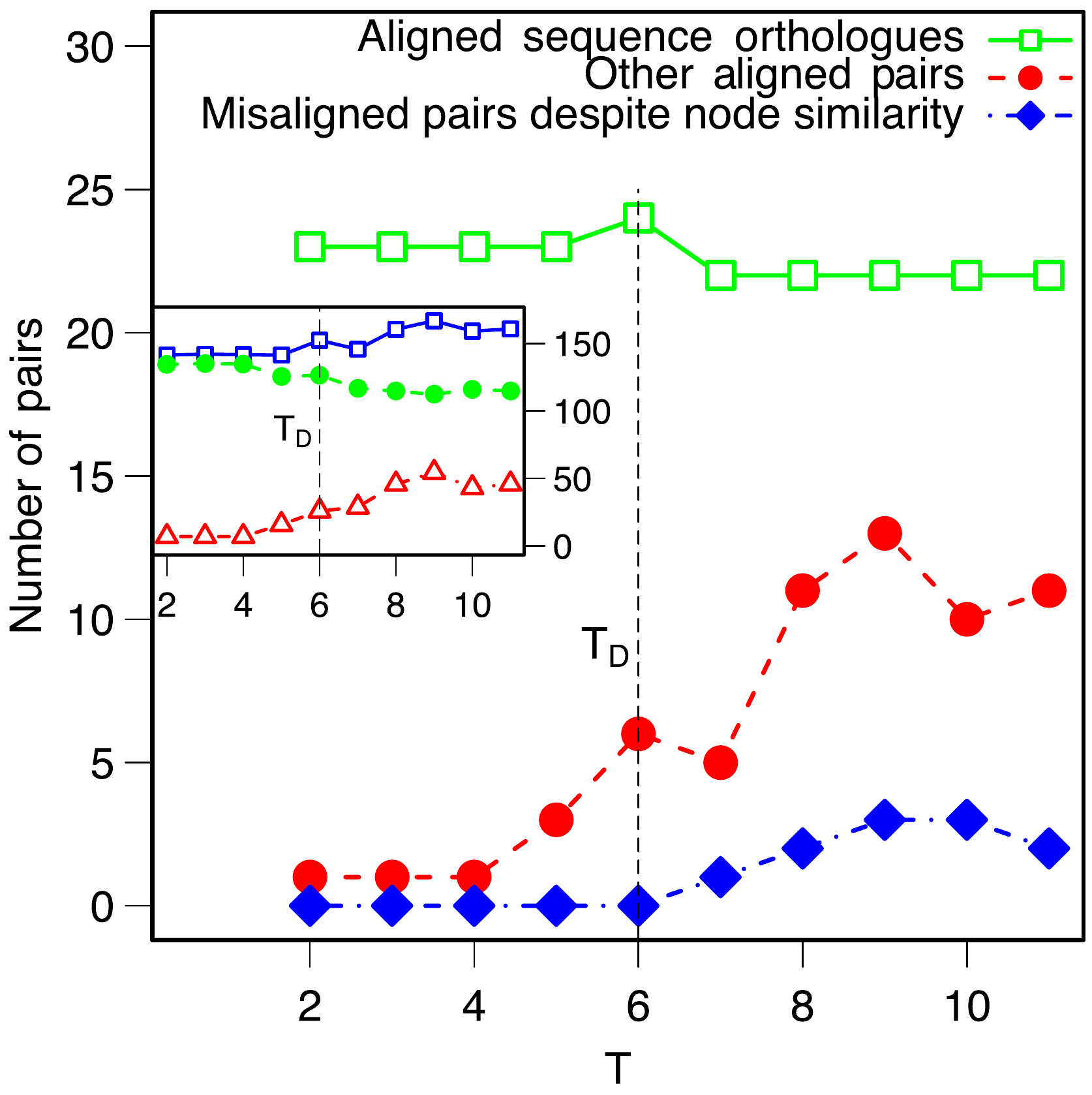}
\caption{{\small {\bf Choice of the temperature $T$ for the real data.} For
  the legend see  Figure~\ref{figure2}. {\bf Main
  figure:} The numbers of aligned pairs of the two herpesviral
  protein interaction networks show similar trends as in the tests
  with artificial data. The threshold temperature is estimated to
  $T_D=6$. {\bf Inset:} Also the observed scores show similar
  dependence on the temperature $T$ as in the case of artificial
  data. By comparison with the random test in the Figure~\ref{figure2}
  we choose the optimal temperature $T = 5$. For the corresponding
  $\rho$-matrix see the Figure~3(b) of the main text.}}
\label{figure3}
\end{figure*}

The optimal temperature for the algorithm run has been estimated from
the Figure~\ref{figure3}.  By comparison of the plot with the
Figure~\ref{figure2} we estimate the value of the transition
temperature $T_D=6$ and run the alignment algorithm with
the schedule defined by $T=5$. In this way, we maximise the number of
aligned pairs while aiming to keep the estimated number of wrongly aligned
pairs negligible. The alignment contains $26$ node pairs out $84$ of KSHV 
and $76$ of VZV (approximately $33\%$).

The list of pairs of Orfs that are present in the resulting alignment
is shown in the Table~\ref{alignedPairs0} together with local scores
for the pairs. The local scores give the contributions of the pair to
the total node and link scores of the alignment. Comparison of the
sequences of the pairs of Orfs which are discussed in the Results
section of the main text are summarised in the
Table~\ref{alignedPairs1}. The comparison of the interaction patterns of
these pairs is summarised in the Table~\ref{alignedPairs2}. 
\begin{table}
  \centering
  \begin{tabular}{|c|c|c|c|}
	\hline
  	{\bf KSHV Orf} & {\bf VZV Orf} & {\bf node score} & {\bf link score} \\
	\hline
	28 & 65 &  3.50 &  6.30 \\
	29b & 42 &  4.30 &  6.14 \\
	67.5 & 25 &  4.20 &  4.57 \\
	23 & 39 & -0.49 &  4.47 \\
	41 & 60 & -0.49 &  4.39 \\
	61 & 19 &  5.41 &  2.00 \\
	60 & 18 &  5.41 &  1.67 \\
	9 & 28 &  5.41 &  0.91 \\
	6 & 29 &  5.41 &  0.35 \\
	25 & 40 &  5.41 &  0.35 \\
	37 & 48 &  5.41 &  0.35 \\
	20 & 35 &  4.66 &  0.35 \\
	29a & 45 &  4.30 &  0.35 \\
	43 & 54 &  5.41 &  0.35 \\
	70 & 13 &  5.41 &  0.35 \\
	8 & 31 &  5.41 &  0.35 \\
	7 & 30 &  5.41 &  0.14 \\
	44 & 55 &  5.41 &  0.14 \\
	19 & 34 &  5.41 &  0.06 \\
	56 & 6 &  5.41 &  0.01 \\
	53 & 9a &  2.49 & -0.08 \\
	17 & 33 &  5.41 & -0.16 \\
	39 & 50 &  5.41 & -0.29 \\
	26 & 41 &  5.21 & -0.29 \\
	46 & 59 &  5.41 & -0.29 \\
	68 & 26 &  5.41 & -0.76 \\
	\hline
\end{tabular} 
  \caption{{\small \textbf{The list of Orfs in the optimal alignment.} 
  The aligned node pairs are ordered according to the value of the link score.}}
  \label{alignedPairs0}
\end{table}

\begin{table}
\centering 
\begin{tabular}{|p{0.45in}|p{0.45in}|p{0.3in}|p{0.45in}p{0.45in}p{0.45in}|p{0.45in}p{0.45in}p{0.45in}p{0.45in}|}
\hline
{\bf KSHV Orf} & {\bf VZV \hspace{0.2in}Orf} & {\bf seq. orth.} & {\bf \textsl{sequenceAlign}} identity (\%) & \hspace{0.5in} length & \hspace{0.5in} score & {\bf \textsl{clustalW}} \hspace{0.2in} identity (\%) & \hspace{0.5in} length & \hspace{0.5in} score & \hspace{0.7in} p-value \\
\hline
67.5    &25     &*      &20.0  &80  &0      &18.4   &76  &57  &0.44 \\
28      &65     &*      &10.8   &102 &0      &10.8   &102 &-31 &0.66 \\
23      &39     &       &---   &--- &---    &17.5   &240 &41  &0.43 \\
41      &60     &       &---   &--- &---    &11.9   &160 &-42 &0.94 \\
\hline
\end{tabular}
\caption{{\small {\bf The sequence similarity of the pairs that are discussed in the Result Section of the main text.} Results of \textsl{sequenceAlign} and \textsl{clustalW} are shown. The p-values for the \textsl{clustalW} results are calculated from the ensemble of randomised sequences as discussed in the main text.}}\label{alignedPairs1}
\end{table}

\begin{table}
\centering 
\begin{tabular}{|p{0.45in}|p{0.45in}|p{0.8in}p{0.8in}|p{0.8in}p{0.8in}|p{0.8in}|}
\hline
{\bf KSHV Orf} & {\bf VZV \hspace{0.2in}Orf} & \mbox{{\bf links in the alignment}} \hspace{0.3in} {\bf KSHV} & \hspace{0.7in} {\bf VZV} & {\bf shared links} & {\bf link score} & {\bf p-value}\\
\hline
67.5   &25    &5    &12  &4   &4.57     &$4\times 10^{-3}$\\
28     &65    &4    &5   &4   &6.30     &$1\times 10^{-3}$\\
23     &39    &4    &4   &3   &4.47     &$2\times 10^{-2}$\\
41     &60    &3    &6   &3   &4.39     &$2\times 10^{-2}$\\
\hline
\end{tabular}
\caption{{\small {\bf Topological similarity of the pairs aligned because of conservation of PIN topology.} The numbers of common links and other links in the aligned subnetwork of the KSHV and VZV network are listed, together with the resulting link score. The p-values are given by equation (\ref{pValue1}).}}
\label{alignedPairs2}
\end{table}

\begin{figure*}	
\centering
{\includegraphics[height=0.4\textwidth]{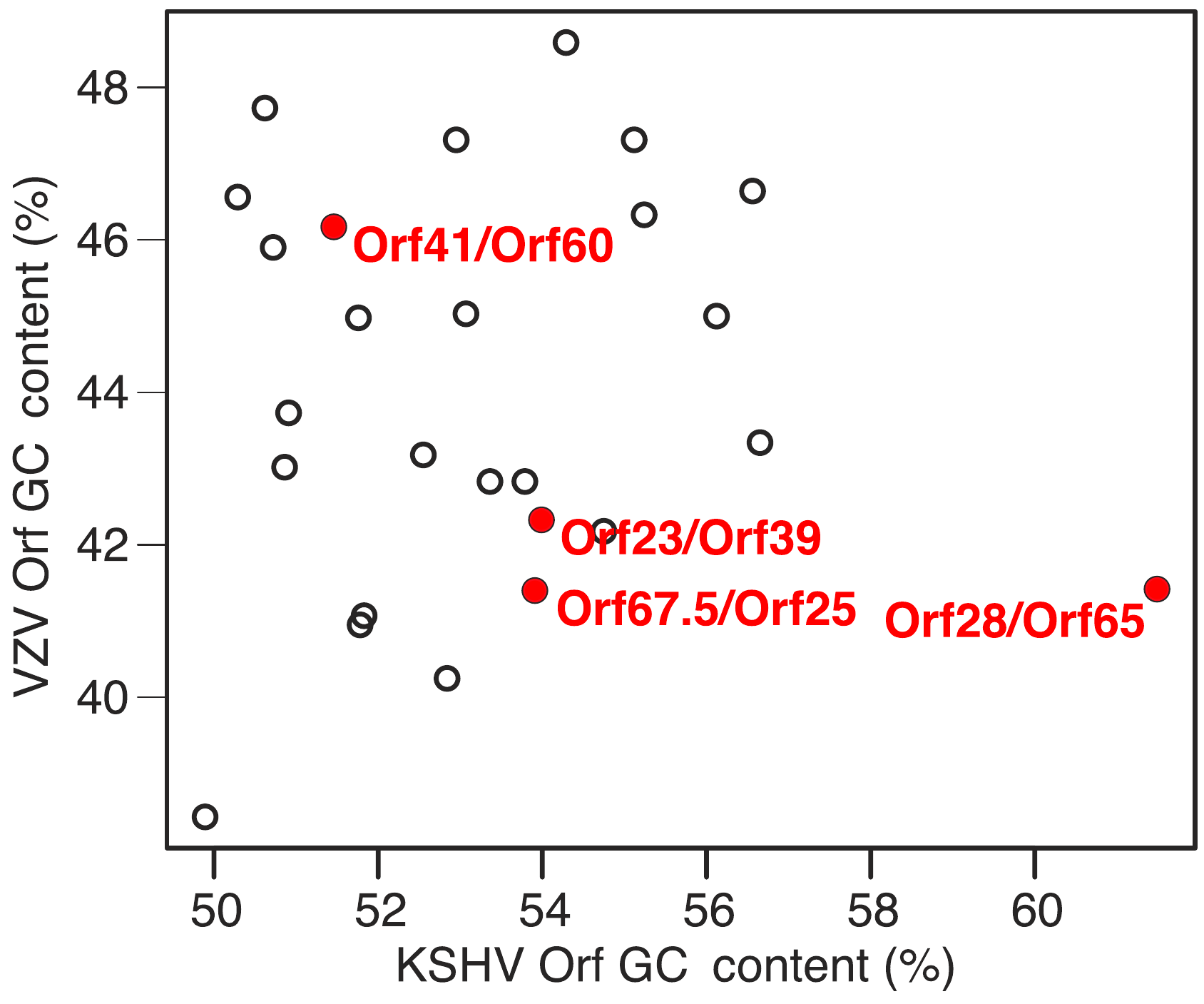}}
\caption{{\small {\bf GC content analysis does not show any correlation.} The
  correlation of GC content has decayed during the independent
  evolution of the two viruses.}}
\label{figure5}
\end{figure*}

Together with other characteristics of the aligned Orfs (the sequence
length and the position in the genome described in the main text), we
compared also the GC content of the aligned pairs. The plot in the
Figure~\ref{figure5} shows that there is no correlation of this
sequence characteristic. The fact that also very closely related
herpesviral Orfs may have very different GC contents has been observed
already by Vl\v{c}ek \emph{et al.} in \cite{vl-etal-1995}.

\subsection{Conservation of the network--aligned Orfs pairs at the sequence level}
The pairwise sequence comparison described in 
Section~\ref{sequencesComparison} have not yielded a significant sequence similarity
for $2$ node pairs aligned solely due to their interaction similarity. 

To further test the possibility of detection of sequence homology homology 
we have searched for multiple sequence alignments of the protein
families to which these Orfs belong. We have extracted the respective
families from the VOCs database~\cite{hi-up-2000}, and compared them
using \textsl{DIALIGN}~\cite{dialign}, \textsl{Parallel PRRN}~\cite{pprrn}, \textsl{MUSCLE}~\cite{muscle}, \textsl{T-COFFEE}~\cite{t-coffee}, \textsl{PSALIGN}~\cite{psalign}, \textsl{SAM-T99}~\cite{sam-t99}, and \textsl{MSA}~\cite{msa}.

For each pair KSHV 67.5/VZV 25, 28/65, 23/39, 41/60 we have selected
from the VOCs database a representative subset of the herpesviral
proteins in the same family (at least ten or all proteins) and
compared these families using the multiple alignment searching
tools. While for the pair 67.5/25 we have found very weak
alignment\footnote{T-Coffee alignment has two stretches of more than 20 aa with $CORE > 3$ (T-Coffee 5.05 EMBL-EBI, default configuration).} of the corresponding families, for the other three pairs we have detected no sequence similarity. This observation further shows
the extend of the evolutionary divergence for the pairs of Orfs.

\begin{figure}
\begin{center}
\includegraphics[width=0.7\textwidth]{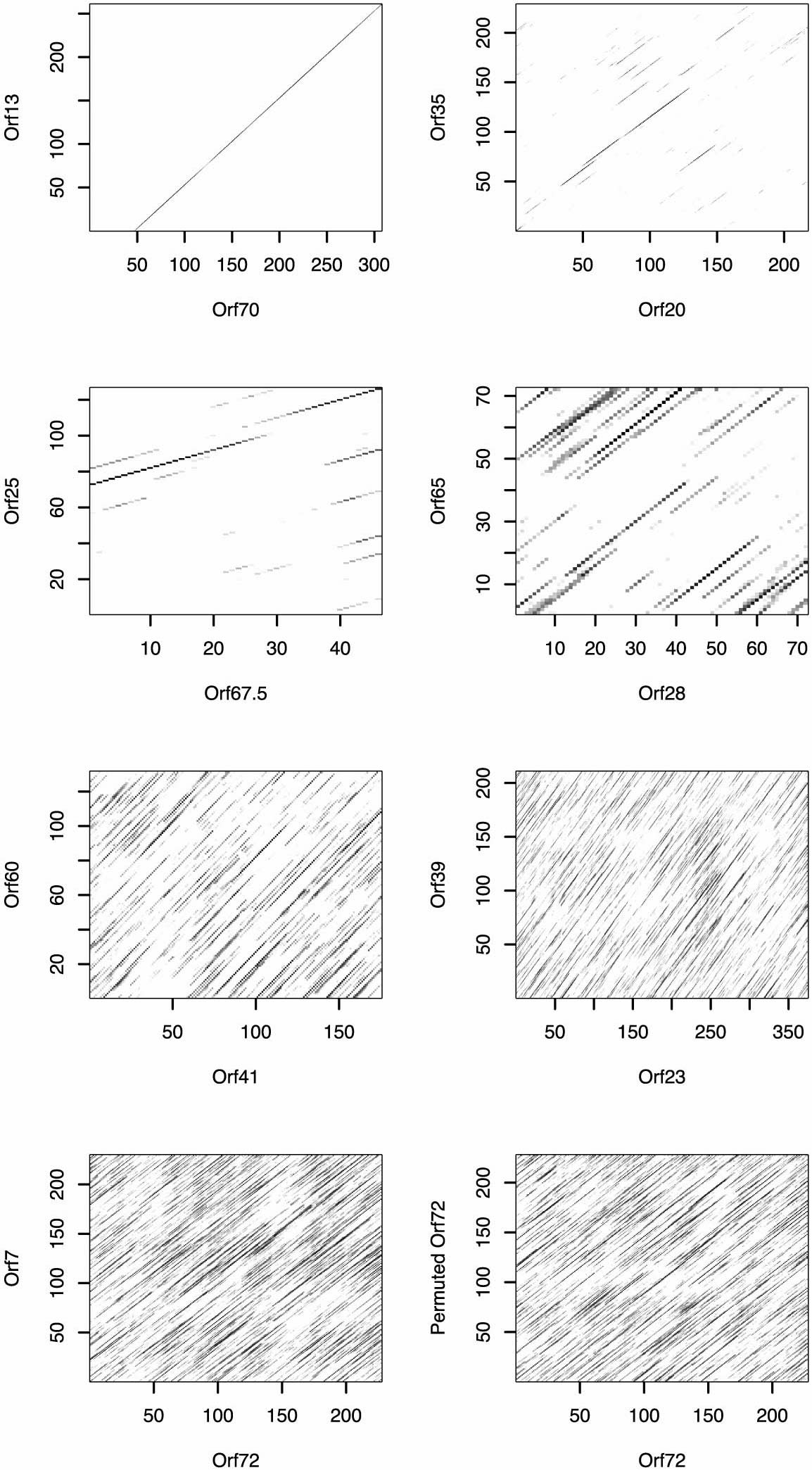}
\caption{{\small {\bf Network alignment allows detection of homologs with poor sequence similarity}. With decreasing level of sequence conservation the dominant diagonal in the dot plot disappears among traces of random alignments. From top--left to bottom--right: Orfs KSHV 70/VZV 13, an almost perfect match; Orfs 35/20, a typical match of sequence homologs, Orfs 67.5/25, the pair aligned due to sequence and network conservation; Orfs 28/65, 41/60 and 23/39, the pairs aligned dominantly or only because of interaction conservation; Orfs 72/7, spurious sequence homologs not aligned by the network alignment; Orf 72/permuted Orf 72, comparison with a random sequence. The sliding window of size 30 has been used for the generation of the dot plots.}}
\label{dotPlots}
\end{center}
\end{figure}

Returning back to the pairwise alignment we have generated the dot
plots for the pairs listed in the Table~\ref{seqOrth}. Here we observe
a very clear pattern: while for the Orfs pairs with a high similarity
the dot plot is dominated by a single diagonal, with increasing
divergence this diagonal disappears among short diagonal lines that
correspond to random alignments, see Figure~\ref{dotPlots}. The
network--aligned pairs show the dot--plot pattern of an intermediate
quality.

\begin{figure}
{\footnotesize
{\normalsize a) {\sl sequenceAlign}}
\begin{verbatim}
Query=  KSHV-BC1-Orf67.5            Length= 80
Sbjct=  VZV-Oka_p-Orf25             Length= 156
Score=  0.2433
logMu  -6.798, logNu 0.006, logAlpha 0.049, 
Matrix: blosum50

Q:                EYAS--------------------------------------------------------
S:                YESENASEHHPELEDVFSENTGDSNPSMGSSDSTRSISGMRARDLITDTDVNLLNIDALE
                    +                                                         

Q:                ----------------DQLLPRDMQILFPTIYCRLNAINYCQYLKTFLVQR--------A
S:                SKYFPADSTFTLSVWFENLIPPEIEAILPTTDAQLNYISFTSRLASVLKHKESNDSEKSA
                                  ++L+P +++ ++PT   +LN I++ + L + L ++        A

Q:                QPAACDHTLVLESKVDTVRQVLRKIVSTDAVFSEA
S:                YVVPCEHSASVTRRRERFAGVMAKFLDLHEILKDA
                      C+H+  +  + +    V+ K +    ++ +A
\end{verbatim}

{\normalsize b) {\sl clustalW}}
\begin{verbatim}
Sequence 1: KSHV-BC1-Orf67.5               80 aa
Sequence 2: VZV-Oka_p-Orf25                156 aa
Alignment Score 57
CLUSTAL W (1.81) multiple sequence alignment

KSHV-BC1-Orf67.5  --------MEYAS----DQLLPRDMQILFPTIYCRLNAINYCQYLKTFLVQRAQP-----
VZV-Oka_p-Orf25   ESKYFPADSTFTLSVWFENLIPPEIEAILPTTDAQLNYISFTSRLASVLKHKESNDSEKS
                            ++     ++L+P +++ ++PT  ++LN I++ + L ++L ++ +      

KSHV-BC1-Orf67.5  ---AACDHTLVLESKVDTVRQVLRKIVSTDAVFSEARARP
VZV-Oka_p-Orf25   AYVVPCEHSASVTRRRERFAGVMAKFLDLHEILKDA----
                     ++C+H+  +  + + +  V+ K+++ + ++++A    
\end{verbatim}
%KSHV-BC1-Orf67.5  ------------------------------------------------------------
%VZV-Oka_p-Orf25   MYESENASEHHPELEDVFSENTGDSNPSMGSSDSTRSISGMRARDLITDTDVNLLNIDAL
%                            ::     ::*:* ::: ::**  .:** *.: . * :.* :: .      
%                     ..*:*:  :  : : .  *: *::. . ::.:*    
}
\caption{{\small {\bf Weak sequence similarity of the KSHV Orf 67.5 and the VZV Orf 25.} Both \textsl{sequenceAlign} \textbf{(a)} and \textsl{clustalW} \textbf{(b)} find alignments with 20\% (18\%) identity over approximately 80 aa. The score of the \textsl{sequenceAlign} alignment is $0.2$, meaning that the random model is almost as likely as the model of evolutionary related sequences. This is also shown by the very high p-value of the \textsl{clustalW} alignment, $p = 0.44$. Such a level of sequence similarity cannot prove homology of the two Orfs, if it is not supported by another method.}}
\label{67.5/25}
\end{figure}

\begin{figure}
{\footnotesize
\begin{verbatim}
Sequence 1: KSHV-BC1-028      102 aa
Sequence 2: VZV-Oka_p-065     102 aa
Alignment Score -31
CLUSTAL W (1.81) multiple sequence alignment

KSHV-BC1-028         MSMTSPSPVTGGMVDGSVLVRMATKPPVIGLITVLFLLVIGACVYCCIRVFLAARLWRAT
VZV-Oka_p-065        MAGQNTMEGEAVALLMEAVVTPRAQPNNTTITAIQPSRSAEKCYYSDSENETADEFLRRI
                     M+  ++    +  +  +++V   ++P    + ++        C Y+  +   A ++ R  

KSHV-BC1-028         PLGRATVAYQVLRTLGPQAGSHAPPTVGIATQEPYRTIYMPD
VZV-Oka_p-065        GKYQHKIYHRKKFCYITLIIVFVFAMTGAAFALGYITSQFVG
                        + ++ ++      +    ++ + +G A    Y T  + +
\end{verbatim}
%                     *:  ..    .  :  ..:*   ::*    : ::        * *.  .   * .: *  
%                        : .: ::      .    .. . .* *    * *  : .
}
\caption{{\small {\bf Homology of the Orfs KSHV Orf28 and VZV Orf65 cannot be inferred from the  sequence comparison only.} The alignment produced by \textsl{sequenceAlign} and \textsl{clustalW} are identical. The p-value of the alignment is $p = 0.66$, though. Another method is needed to support the homology.}}
\label{28/65}
\end{figure}

\begin{figure}
{\footnotesize
\begin{verbatim}
Sequence 1: KSHV-BC1-023         404 aa
Sequence 2: VZV-Oka_p-039        240 aa
Alignment Score 41
CLUSTAL W (1.81) multiple sequence alignment

KSHV-BC1-023         MLRVPDVKASLVEGAARLSTGERVFHVLTSPAVAAMVGVSNPEVPMPLLFEKFGTPDSST
VZV-Oka_p-039        ----------------------------------------------------------MN
                                                                                +

KSHV-BC1-023         LPLYAARHPELSLLRIMLSPHPYALRSHLCVGEETASLGVYLHSKPVVRGHEFEDTQILP
VZV-Oka_p-039        PPQARVSEQTKDLLSVMVNQHP--------------------------------------
                      P   + +   +LL +M++ HP                                      

KSHV-BC1-023         ECRLAITSDQSYTNFKIIDLPAGCRRVPIHAANKRVVIDEAANRIKVFDPESPLPRHPIT
VZV-Oka_p-039        --------------------------------------EEDAKVCKSSDNSPLYNTMVML
                                                           +E A+  K  D ++      + 

KSHV-BC1-023         PRAGQTRSILKHNIAQVCERDIVSLNTDNEAASMFYMIGLRRPRLGESPVCDFNTVTIME
VZV-Oka_p-039        SYGGDTDLLLSS----ACTRTSTVNRSAFTQHSVFYIIST----VLIQPICCIFFFFYYK
                     + +G+T  +L+     +C R  +  ++     S+FY+I+     +  +P+C +  +   +
          
KSHV-BC1-023         RANNSITFLPKLKLNRLQHLFLKHVLLRSMGLENIVSCFSSLYGAELAPAKTHEREFFGA
VZV-Oka_p-039        ATRCMLLFTAGLLLTILHHFRLIIMLL----------CVYRNIRSDLLPLSTSQQLLLGI
                      ++  + F + L L+ L+H+ L  +LL          C+     ++L P +T ++ ++G 
                                
KSHV-BC1-023         LLERLKRRVEDAVFCLNTIEDFPFREPIRQPPDCSKVLIEAMEKYFMMCSPKDRQSAAWL
VZV-Oka_p-039        IVVTRT-----MLFCITAYYTLFIDTRVFFLITGHLQSEVIFPDSVSKILPVSWGPSPAV
                     ++   +      +FC+++   + +   +             + + +    P +  +++ +

KSHV-BC1-023         GAGVVELICDGNPLSEVLGFLAKYMPIQKECTGNLLKIYALLTV
VZV-Oka_p-039        LLVMAAVIYAMDCLVDTVSFIG-----PRVWVRVMLKTSISF--
                        ++ +I   + L ++++F++      +  +  +LK    +  
\end{verbatim}
%                                                                                .
%                      *   . .   .** :*:. **                                      
%                                                           :* *:  *  * ..      : 
%                     . .*:*  :*.     .* *  .  .:     *:**:*.     :  .*:* :  .   :
%                      :.  : * . * *. *:*: *  :**          *.     ::* * .* :: ::* 
%                     ::   .      :**:.:   : :   :             : . .    * .  .:. :
%                        :. :*   : * :.:.*:.      :  .  :**    :  
}
\caption{{\small {\bf Homology of KSHV Orf 23 and VZV Orf 39 cannot be inferred from sequence similarity.}
The \textsl{sequenceAlign} does not find any nontrivial alignment and the \textsl{clustalW} alignment has the p-value equal to $0.43$. Another method is needed to support the homology.}}
\label{23/39}
\end{figure}

\begin{figure}
{\footnotesize
\begin{verbatim}
Sequence 1: KSHV-BC1-041         205 aa
Sequence 2: VZV-Oka_p-060        160 aa
Alignment Score -42
CLUSTAL W (1.81) multiple sequence alignment

KSHV-BC1-041         MAGFTLKGGTSGDLVFSSHANLLFSTSMGYFLHAGSPRSTAGTGGEPNPRHITGPDTEGN
VZV-Oka_p-060        ---------MASHKWLLQMIVFLKTITIAYCLHLQDDTPLFFGAKPLSDVSLIITEPCVS
                               +++  + +   +L + +++Y LH  +  +    +   +   +  +++  +

KSHV-BC1-041         GEHRNSPNLCGFVTWLQSLTTCIERALNMPPDTSWLQLIEEVIPLYFHRRRQTSFWLIPL
VZV-Oka_p-060        SVYEAWDYAAPPVSNLSEALSGIVVKTKCP--------VPEVILWFKDK--QMAYWTNPY
                     + ++     +  V+ L++  + I    + P        + EVI  + ++  Q ++W  P 

KSHV-BC1-041         SHCEGIPVCPPLPFDCLAPRLFIVTKSGPMCYRAGFSLPVDVNYLFYLEQTLKAVRQVSP
VZV-Oka_p-060        VTLKGLTQSVGEEHKSGDIRDALLDALSGVWVDS------------------------TP
                        +G++ +    +++   R  ++   + +   +                        +P 

KSHV-BC1-041         QEHNPQDAKEMTLQLEAWTRLLSLF
VZV-Oka_p-060        SSTNIPENGCVWGADRLFQRVCQ--
                     ++ N  +   +    + + R+ +  
\end{verbatim}
%                               :..  : .   :* : ::.* **  .  .    .   .   :  .:.  .
%                     . :.     .  *: *..  : *    : *        : ***  : .:  * ::*  * 
%                        :*:. .    ...   *  ::   . :   :                        :*
%                     .. *  :   :    . : *: .  
}
\caption{{\small {\bf \textsl{ClustalW} alignment of the KSHV Orf 41 and VZV Orf 60 is not statistically significant.} Its p-value is 0.94. \textsl{SequenceAlign} does not find any nontrivial alignment.}}
\label{41/60}
\end{figure}

\subsection{Conserved links typically connect alike Orfs}
To examine the relationship between function of the proteins and the conservation of links among them, we analyse the likeliness of the conservation of the links among the proteins with similar functions and of the conservation of the links between the proteins with dissimilar functions. 

First we test if the conserved links are more likely to connect alike proteins. To do so, we group the Orfs to two functional classes: the protein belongs either to the class of `structure--related' proteins (classes: capsid/core protein, membrane/glycoprotein, virion protein, virion assembly) or to the class of `information--processing' proteins (DNA replication, gene expression regulation, nucleotide repair/metabolism, host--virus interaction). We calculate the frequencies $p(f_i,f_j)$ of functional annotations $f_i$, $f_j$ of adjacent Orfs $i$ and $j$ in the subgraph containing all sequence homologs (resp. all network--aligned Orfs),
\begin{equation}
p(g,	 h) = \frac{\sum_{links} \delta(f_i = g) \delta(f_j = h)}
			   {\sum_{links} 1}.
\end{equation}
We evaluate this sum separately for all the links in the subgraph ($p_A$), the conserved links only ($p_M$) and for the nonconserved (mismatching) links in the subgraph ($p_{MM}$).

Then we calculate the mutual information as the measure of the influence of the functional annotation of the adjacent Orfs on the link state,
\begin{equation}
I = \sum_g \sum_h p(g,h) \ln \frac{p(g,h)}{p(g)p(h)},
\end{equation}
where $p(g)$ is the marginal $p(g) = \sum_h p(g,h)$. Keeping the subscripts we find: $I_A = 0.0014$ for the subgraph of sequence homologs ($0.0092$ for the alignment subgraph), $I_M = 0.0743$ ($0.1178$), and $I_{MM} = 1\times 10^{-6}$ ($0.0001$). Clearly, the largest mutual information of functional annotation of adjacent Orfs is among nodes connected by matching links (by a factor of 100 or more). We have evaluated also the frequency tables for the complete protein interaction networks, $p_K$ and $p_V$. Not surprisingly, the mutual information is small for these graphs $I_K = 0.0030$, and $I_V = 0.0052$.

To estimate the p-value of such a mutual information we have reshuffled the positions of the conserved links randomly and we have evaluated the mutual information $I_M$ for such randomised graphs. The probability of finding equal or better mutual information $I_M$ in an ensemble of $10^5$ graphs generated in this way has been taken as the p-value. The estimates are $0.12$ for the subgraph of sequence homologs and $0.05$ for the alignment subgraph.

Secondly, we test if the links between similar proteins are more likely to be conserved. Taking the subgraph of the sequence homologs as the basis of our analysis, we create the matrix $n_F(a,b)$ defined in the following way: $n_F(0,0)$ is the number of pairs of the alike Orfs between which there is a link in neither species; $n_F(1,0)$ is the number of the pairs of the alike Orfs that are connected by a link in KSHV solely; $n_F(0,1)$ the same for VZV; and $n_F(1,1)$ is the number of conserved links between alike Orfs. We create the second matrix $n_D$ defined similarly for the pairs of Orfs with unlike functional annotation.

Then we define the conservation ratio $p_F$ as the ratio of the number of conserved links and the total number of links
\begin{equation}
c_F = n_F(1,1) / ( n_F(0,1) + n_F(1,0) + n_F(1,1) ).
\end{equation}
In the same way we define $c_D$ for the links between unlike Orfs.

A rough estimate of the odds in the link conservation can be expressed
as the ratio of the two conservation ratios
\begin{equation}
C_F = c_F / c_D.
\end{equation}
Its value is $1.62$, that is the links between alike Orfs are $62\%$
more likely to be conserved than the links between unlike Orfs ($c_F =
0.15$, $c_D = 0.09$). The p-value evaluated over the same graph with a
randomised annotation list is $0.47$.

More refined estimate of the odds which takes in consideration also
the link statistics of the two networks uses mutual information, $I_F$
and $I_D$. The mutual information expresses the level of correlation
of the presence of the link in the two networks. If we normalise the
frequencies $n_F$, $p_F(a,b) = n_F(a,b) / \sum_{c,d} n_F(c,d)$, we may
write the mutual information as
\begin{equation}
I_F = \sum_{a,b} p_F(a,b) \ln \frac{p_F(a,b)}{p_F^A(a)p_F^B(b)},
\end{equation}
where the marginals are defined as $p_F^A(a) = \sum_b p_F(a,b)$ and
$p_F^B(a) = \sum_a p_F(a,b)$. In the same way we define the mutual
information $I_D$ for the unlike Orfs pairs. For the subgraph of the
sequence homologs the mutual information reads $I_F = 0.049$, $I_D =
0.005$. Defining the final information odds, $D_F = I_F / I_D$, we get
$D_F = 9.58$ with the p-value 0.13 (the same test as for $C_F$).

\end{document}